%% file: ICNTmaster_v2.tex
\documentclass[epj]{svjour}

\usepackage{amsmath}	
\usepackage{color}
\usepackage{mathtools, cuted}
\usepackage{hyperref}
\pdfoutput=1
\newcommand{\cev}[1]{\reflectbox{\ensuremath{\vec{\reflectbox{\ensuremath{#1}}}}}}

\title{Toward a complete theory for predicting inclusive deuteron breakup away from stability}

\author{G. Potel \inst{1} \email{potel@nscl.msu.edu} \and G. Perdikakis \inst{1,}\inst{2,}\inst{3,}  \email{perdi1g@cmich.edu} \and  B. V. Carlson \inst{4} \email{brett@ita.br} \and M. C. Atkinson \inst{5} \and  W. H. Dickhoff \inst{5} \and J. E. Escher \inst{6} \and M. S. Hussein \inst{4,}\inst{7,}\inst{8} \and J. Lei \inst{9,}\inst{10}\thanks{Present address} \and W. Li \inst{1} \and    A. O. Macchiavelli \inst{11} \and A. M. Moro \inst{9} \and  F. M. Nunes \inst{1,} \inst{12} \and S. D. Pain \inst{13} \and J. Rotureau \inst{1}}
\institute{Facility for Rare Isotope Beams, Michigan State University, East Lansing, MI 48824, USA \and Department of Physics, Central Michigan University, Mt. Pleasant, MI 48859, USA \and Joint Institute for Nuclear Astrophysics--Center for the Evolution of the Elements, East Lansing, MI 48824,USA \and Instituto Tecnol\'ogico de Aeron\'autica, DCTA, 12.228-900 S\~{a}o Jos\'e dos Campos, SP, Brazil \and Department of Physics, Washington University, St. Louis, MO 63130, USA \and  Lawrence Livermore National Laboratory, Livermore, California 94550, USA \and Departamento de F\'isica Matem\'atica, Instituto de F\'isica, Universidade de S\~ao Paulo, C.P. 66318, 05314-970, S\~ao Paulo, SP, Brazil \and Instituto de Estudos Avan\c{c}ados, Universidade de S\~ao Paulo, C.P. 72012, 05508-970, S\~ao Paulo, SP, Brazil \and Departamento de FAMN, Universidad de Sevilla, Apartado 1065, 41080 Sevilla, Spain \and Institute of Nuclear and Particle Physics, Ohio University, Athens, Ohio 45701, USA \and Nuclear Science Division, Lawrence Berkeley National Laboratory, Berkeley, California 94720, USA \and Department of Physics and Astronomy, Michigan State University, East Lansing, Michigan 48824-1321, USA \and Physics Division, Oak Ridge National Laboratory, Oak Ridge, Tennessee 37831, USA }

\begin{abstract}
 {We present  an  account of the current status of the theoretical treatment of inclusive $(d,p)$ reactions in the breakup--fusion formalism, pointing to some applications and making the connection with current experimental capabilities. Three  independent implementations  of the reaction formalism have been recently developed, making use of different numerical strategies. The  codes also originally relied on two different but equivalent representations, namely the prior (Udagawa--Tamura, UT) and the post (Ichimura--Austern--Vincent, IAV) representations.
  The different implementations have been  benchmarked, and then applied to the Ca isotopic chain. The neutron--Ca propagator is described in the Dispersive Optical Model (DOM) framework, and the interplay between elastic breakup (EB) and non--elastic breakup (NEB) is studied for three Ca isotopes at two different bombarding energies. The accuracy of the description of different reaction observables is assessed by comparing with experimental data of $(d,p)$ on $^{40,48}$Ca. We discuss the predictions of the model for the extreme case of an isotope ($^{60}$Ca) currently unavailable experimentally, though possibly available in future facilities (nominally within production reach at FRIB). We explore the use of $(d,p)$ reactions as surrogates for $(n,\gamma)$ processes, by using the formalism to describe the compound nucleus formation in a $(d,p\gamma)$ reaction as a function of excitation energy, spin, and parity. The subsequent decay is then  computed within a Hauser--Feshbach formalism. Comparisons between the $(d,p\gamma)$ and $(n,\gamma)$ induced gamma decay spectra are discussed to inform efforts to infer neutron captures from $(d,p\gamma)$ reactions. Finally, we identify areas of opportunity for future developments, and discuss a possible path toward a predictive reaction theory.
}
\end{abstract}

\begin{document}
\authorrunning{G. Potel \textit{et al.}}
\titlerunning{Inclusive deuteron breakup}
\maketitle

\section{Introduction and Motivation}
\input{Introduction}

\section{Theory}\label{Reaction_Theory}
\subsection{Reaction  formalism}\label{Reaction}
\input{Formalism}
\subsection{Details of the Surrogate technique and the formation of the compound system}\label{surrogates}
\input{Application_Surrogates}
\subsection{Compound system decay}\label{decay}

\input{Compound_decay}

%
\subsection{Nucleon--nucleus interaction}\label{Structure}
\input{Optical_potentials_related_quantities}
\section{Results}\label{results}
\input{results}

%%\input{dp_dn_to_bound}
%
%%\input{dp_dn_to_resonances_continuum}
%
%%\input{Gamma_decay}
\input{Results_gamma_decay}

%
%\input{Sensitivity_uncertainties}
%
\section{Conclusions} \label{Discussion}
\input{Discussion}

%%%%%%%%%%%%%%%%%%%%%%%%
\clearpage

\providecommand{\newblock}{}

%\bibliography{references}% Produces the bibliography via BibTeX.
\end{document}

%% file: Introduction.tex
Nuclear reactions provide an important source of structure information. Deuteron induced reactions $d+A$ in particular have played a prominent role in our field. From the experimental perspective they are attractive because these reactions can be studied both in direct kinematics (the deuteron being the beam) and in inverse kinematics (where the target is deuterated), and  
thus allow one to probe a wide spectrum of phenomena. 
Among the various reaction channels of interest, elastic and inelastic scattering, as well as neutron transfer have been extensively explored since the sixties. A much smaller body of work can be found in the literature for $(d,n)$ reactions due to the increased difficulty in measuring the neutron. Given the loosely bound nature of the deuteron,  elastic breakup $A(d,pn)A$, where both neutrons and protons fly away from the target, leaving the target in its ground state, is an important channel. In addition, the deuteron induces a variety of non-elastic processes, including processes leading to the formation of a compound nucleus, which are also of interest for many applications. 

Recently there has been an upsurge of interest in applying the theory of inclusive  breakup to  $(d, p)$ reactions. The motivation for such research activity is the potential application of the theory to extract neutron or proton capture cross sections. These can be used to address questions on the origin and synthesis of the elements in the cosmos (\cite{Arcones:17,Carlson:17b}). Of course, for an answer to these astrophysical questions nuclear physics information has to be included in the modeling of the various astrophysical environments of interest. In particular, for all the  corresponding stellar conditions, reliable  constrained thermonuclear reaction rates have to be available for the variety of nuclear reaction mechanisms accessible to the system at stellar temperatures, and for the vast number of participating nuclei. Two types of thermonuclear processes are predominantly relevant to these environments, the explosive hydrogen burning in Type I X-ray bursters and Novae mediated by proton capture $($p,$\gamma)$ reactions, and the neutron capture processes in Supernovae and Compact Object Mergers.

Factors such as the very low reaction cross sections at the astrophysical temperature of interest, the unavailability of neutron targets or/and the unavailability of the more unstable isotopes in adequate beam intensities, limit the degree to which such reactions can be studied directly experimentally.  Consequently, experimental efforts have concentrated on indirect ways to either constrain or deduce the required reaction rates. $(d,p)$  reactions have been used to probe  in an indirect way these $(n,\gamma$) reactions using reaction theory to interpret the results and extract the needed reaction rates. The consistent treatment of the competing reaction mechanisms in deuteron-induced reactions and the application of a self-consistent theoretical treatment for bound and unbound states near and away from stability remains a challenge for nuclear reaction studies.

Similar experimental challenges exist for direct studies of neutron capture reactions on radioactive nuclei present in the fuel or the waste of nuclear reactors. It has been demonstrated that problems such as the effective transmutation of nuclear waste in accelerator driven systems or in fast breeder reactors can be studied using deuteron--induced reactions (for a recent example in the case of $^{238}$U see \cite{Ducasse:16}). 
Both types of problems (low reaction rates and unavailability of $n$ targets) can be addressed in an attractive application of deuteron-induced reactions by using inverse-kinematics experiments with exotic beams. \subsection{Theory challenges}
From a theoretical perspective, deuteron induced reactions are attractive, mainly due to the simple structure of the deuteron: the complex many-body problem $d+A$ can be approximately treated as a three-body problem $n+p+A$. Even though this presents a dramatic simplification of the problem, challenges remain.  While the solution to the  few-body scattering problem is well under control for  few-nucleon systems, it still poses difficulties for deuteron induced reactions on nuclear targets, particularly due to the Coulomb interaction.  The reduction from the many--body problem to a three--body description, requires the introduction of effective nucleon--nucleus interactions that are not well studied. Microscopically, these interactions are  nonlocal, which for many implementations increases the numerical difficulty of the problem. They should also be dispersive, connecting the bound states of the $A+N$ system to the scattering states. 
This brings us to one of the underlying science questions of rare isotope physics, i.e., how the properties of protons and neutrons 
in the nucleus change from the valley of stability to the respective drip lines.

Elastic nucleon scattering has traditionally provided insights into this question for stable nuclei by clarifying how a 
nucleon experiences its propagation through the nucleus at positive energy.
This experience is usually represented in terms of an energy-dependent complex potential, the optical potential.
A theoretical framework for this potential was developed by Feshbach~\cite{Feshbach:58,Feshbach:62} employing projection techniques.
A related connection between elastic-scattering data and nucleon propagation was established by Bell and Squires~\cite{Bell:59} demonstrating that the nucleon elastic-scattering $\mathcal{T}$-matrix is equivalent to the reducible self-energy obtained by iterating the irreducible one to all orders with the free nucleon propagator~\cite{Villars67,BlaizotR86,Dickhoff08}.
This provides a more flexible approach in the present context since both reaction and structure information are simultaneously addressed unlike the Feshbach projection formulation which only emphasizes the elastic scattering aspects of the optical potential~\cite{Feshbach:58,Feshbach:62}. Traditionally, positive energy nucleons are described by fitted optical potentials mostly in local form~\cite{Varner91,Koning:03}.
Bound nucleons are usually analyzed with static potentials that lead to an independent-particle model (IPM) modified by the interaction between valence nucleons as \textit{e.g.} in traditional shell-model calculations~\cite{Brown01,Caurier05}.
The link between nuclear reactions and nuclear structure is provided by considering these potentials as representing  different energy domains of one underlying nucleon self-energy.
The seminal work of Mahaux and Sartor emphasized the link between these traditionally separate fields in nuclear physics~\cite{Mahaux86,Mahaux:91}.
Finally, one important question is how does one treat the inelastic processes without giving way to  uncontrolled approximations. It is the goal of our community to develop a predictive (yet feasible) theory that can handle the large array of reaction mechanisms, with quantifiable uncertainties.

One of the first attempts to give a simple and yet semiquantitative measure of the cross section for the observation of one of the fragments while allowing the other fragment to interact with the target, the inclusive non-elastic breakup, was supplied by Serber \cite{Serber:47}. The Serber formula is, in a nutshell, the product of the square of the Fourier transform of the ground state wave function of the projectile and the total reaction cross section of the unobserved fragment. The singles spectrum, observed by experiment, is the sum of the Serber cross section plus the diffractive, elastic breakup cross section. A more rigorous formulation of the breakup process using the theory of direct reactions, resulted in an inclusive breakup cross section which maintains the essentials of the Serber formula, but pays due attention to distortion effects on the observed fragment and the internal motion of the two fragments inside the projectile on their scattering and reaction mechanisms \cite{Baur:76,Kerman:79,Udagawa:81,Baur:84,Ichimura:85,Hussein:85}. The formal separation between elastic breakup and inclusive non-elastic breakup was made possible through the formal developments advanced in \cite{KI1982,Hussein:84}.  The above theories, based on the Distorted Wave Born Approximation (DWBA), were discussed in detail in \cite{Austern:87}, and an exact three-body description of the breakup process was supplied. One issue highlighted in \cite{Austern:87}, was the difference between the post DWBA theory of Ichimura, Austern, and Vincent (IAV) \cite{Ichimura:85}, and the prior DWBA theory of  Udagawa and Tamura (UT) \cite{Udagawa:81}. It was concluded that the UT theory \cite{Udagawa:81} supplies only the two--step process of elastic breakup followed by the capture of the interacting fragment by the target. The more comprehensive IAV theory contains other non-elastic processes, which are accounted for by the Hussein McVoy (HM) theory. A formal demonstration of the inclusiveness of the IAV cross section was given by \cite{Ichimura:90,HFM1990}.

In Ref. \cite{Potel:15b} the UT formula based on the prior form of the DWBA, was employed and an attempt was made to give theoretical foundation for the use of $(d,p)$ reactions in the  Surrogate method \cite{Escher:12rmp}, while Refs. \cite{Lei:15,Lei:15b,Carlson:15} employed the post IAV formulation. The latter was also employed by \cite{Ducasse:16} for the $^{238}$U fast breeder reactor application. The question that arises naturally, is how are these methods related? Refs. \cite{Ichimura:90,HFM1990} derived the relation between the IAV and the UT and HM theories and this has been tested in \cite{Lei:15b}. In this paper we supply further insight into the different approaches to $(d,p)$ reactions, and point out future developments and applications. We also supply a more detailed derivation of the Surrogate cross section.

 \subsection{Experimental context}
 Deuteron-induced reactions have been studied experimentally for decades in normal kinematics utilizing deuteron beams, which can be delivered by relatively modest accelerator facilities, and stable targets. Over the last two decades or so, significant effort has been extended toward the measurement of deuteron-induced reactions in inverse kinematics, in order to increase its applicability to all nuclei that can be made into beams, stable or unstable. 
 
Inverse kinematic measurements introduce a number of experimental challenges over stable-beam measurements, including kinematic compression, a large dynamic range of the energies of light-ion ejectiles (requiring high position resolution), increased target-thickness contributions to resolution. In addition, for radioactive beams, issues regarding beam intensity, emittance and the time structure of some radioactive beams need be addressed. Essentially, two solutions have been identified pertaining to these issues. The optimization of charged particle detection and resolution, or the supplementation of charged particle detection with $\gamma$--ray measurements that are much less impacted by the challenges of inverse kinematics.
 
 The Oak Ridge Rutgers University Barrel Array (ORRUBA) was developed in 2004-2005 as the first large silicon array in the US dedicated to such inverse kinematic measurements, with a primary focus on the measurement of $(d,p)$ reactions on fission fragments. The array is based upon resistive strip detector technology to provide high resolution in angle and position of detection complemented by a large angular coverage around the target (hence the barrel geometry) and using electronics able to cope with the high dynamic range of light particle energies. The array has been used for a series of transfer measurements using ISOL beams at the HRIBF over the 2006 to 2014 period.
 
 An alternative  solution to the inverse kinematics problems and in particular to the effects of kinematic compression on the particle spectra and the need for large angular coverage is the Helical Orbit Spectrometer (HELIOS) device \cite{Lighthall:10}. HELIOS enables the reactions to be performed in a uniform magnetic field, the axis of which is aligned with the beam axis. In this environment, the light-ion ejectiles undergo helical orbits in the magnetic field, ultimately returning to the beam axis from which they were emitted. By measuring their energy, and time and location of return to the beam axis in a silicon array, it is possible to obtain particle identification and a measure of the particle's energy in the center-of-mass system, thereby bypassing the effects of kinematic compression. 
 
% However, some of the issues pertaining to inverse kinematic measurements with radioactive beams, such as target thickness effects and beam emittance, are not addressed by this approach, making solenoidal spectrometers particularly well suited to cases were the beam emmittance is good, and the beam intensity is sufficiently high to allow the use of thin ($\sim 100 \mu$g/cm$^2$) target thicknesses.
 
 The approach of measuring $\gamma$ rays can provide a solution to the target thickness and energy resolution problems in radioactive beam experiments. $\gamma$ ray spectroscopy is less impacted by the beam quality and target thickness. There is an associated efficiency loss to $\gamma$--ray measurements compared with particle spectroscopy (typically about a factor of ten for larger Ge arrays) much of which can be made up with increased target thickness. At low beam energies (e.g. below about 10 MeV/A), Doppler correction using coarse segmentation of coaxial Ge or Clover detectors can be sufficient to significantly exceed the state-of-the-art particle energy resolution. At higher beam energies, the Doppler broadening becomes more significant, requiring higher-spatial resolution for the $\gamma$--ray interaction. To this end, the $\gamma$--Ray Energy Tracking In-beam Nuclear Array (GRETINA) has been developed \cite{Paschalis201344}; this instrument is already revolutionizing the resolution obtainable by transfer reactions in inverse kinematics with fast beams \cite{Kankainen:16}. GRETINA is a first-stage low angular coverage development predecessor of the 4$ \pi $ covering $ \gamma $ ray Energy Tracking Array (GRETA) which is currently under development and will be the workhorse for $ \gamma $-ray measurements at 30-45 MeV/A beam energies with FRIB.
 
 Combining $ \gamma $ ray and particle detection, various Si particle detector couplings to Ge $ \gamma $ ray arrays have been developed over the past decade, (e.g. TIARA, SHARC+TIGRESS, T\_REX+Miniball) to reap the benefits of both solutions. More recently, the Gammasphere and ORRUBA: Dual
 Detectors for Experimental Structure Studies (GODDESS) detector system has been developed, benefiting from the large diameter of Gammasphere to combine the large acceptance of ORRUBA detectors with $\gamma$ ray detection. The coupling of the HELIOS Spectrometer with a $ \gamma $ ray detector array (APOLLO) has also been achieved, albeit this setup has a lower intrinsic  $ \gamma $ ray energy resolution and is mostly aimed at the study of statistical decays of compound nuclei.  
 
% Typically, these systems using a compact Si array (with a typical target-detector distance of a few cm at 90$\circ$ in order to maximize efficiency for gamma-ray detection using a close-packed Ge array. Unusually, the large internal diameter of Gammasphere array of HPGe detectors allows a much larger Si array to be implemented, which is better suited to high-resolution charged particle measurement. Though ORRUBA was developed originally as a stand-alone charged-particle detector, more recently it has been coupled to Gammasphere as part of GODDESS (Gammasphere ORRUBA: Dual Detectors for Experimental Structure Studies). The coupled system provides charged particle detection over 15$^\circ$ to 165$^\circ$ polar angle with 1$^\circ$ resolution in polar angle and $\sim$30 keV energy resolution, along with gamma-ray detection in the $\sim$ 100 HPGe detectors of Gammasphere. The first experiments were performed with GODDESS in July-September 2015.
 Following these developments, there is an ongoing effort to develop a HELIOS-like spectrometer at FRIB and to couple ORRUBA and GRETINA. The latter one will enable high-resolution charged-particle measurement with $\gamma$ rays with sufficient--resolution Doppler reconstruction to match intrinsic Ge resolution. The new developments are expected to enable high-resolution charge particle, $ \gamma $ ray and particle-$ \gamma $ ray measurements with fast beams in the FRIB era.
 
% Maybe something about the challenges of elastics measurements in inverse kin?
 
 %In recent years, much progress has been made to predict many of these inputs using microscopic theories.  To achieve reliable reaction predictions, it is important to complement these predictions with experimental data. Measurements of neutron resonance spacings, for instance, provide stringent constraints on level densities and $\gamma$-induced reactions provide useful information on $\gamma$-ray strength functions.  In addition, indirect methods, such as the surrogate \textcolor{red}{[reference??]} and Oslo \textcolor{red}{[reference???]} methods, aim at extracting relevant information from observing decays of compound nuclei produced by alternative reactions (e.g. from inelastic scattering, transfer reactions, or $\beta$-decay)~\cite{Escher:16a}. 
 %In turn, when the decay of a compound nucleus is well-characterized, it is possible to gain insights into the processes that formed the CN.  For example, the decay of a given compound nucleus depends on its angular-momentum and parity distribution prior to decay, as will be discussed for the case of $^{41}$Ca below.
 
% In this paper we supply further insight into the inclusive breakup--fusion theoretical approach to $(d, p)$  reactions, and point out future developments and applications to, e.g.,  (p, d) reactions. 

\subsection{Latest numerical developments}
Recently, three groups from the University of Sevilla (US, \cite{Lei:15}), Michigan State University and Lawrence Livermore National Lab (MSU/LLNL, \cite{Potel:15b}) and Instituto Tecnol\'ogico de Aeron\'autica (ITA, \cite{Carlson:15}) have developed three different numerical  implementations of the formalism to be described in Sect. \ref{Reaction}. 
Although the physics behind the three codes is the same, the implementations rely on different  strategies. Originally, the MSU/LLNL code was the only one using the prior representation and including energies below the neutron emission threshold. In a recent paper though, the Seville code has been extended to negative  energies \cite{Lei:17}, and included also the calculation in  the prior representation. Lei and Moro were thus able to explicitly verify the post--prior equivalence \cite{Lei:15b}.  On the other hand, the ITA code relies on a zero--range approximation for the proton--neutron interaction, being thus much faster and suitable for extensive calculations, such as the ones needed in large, multi--purpose transport codes. It is now possible to make  a direct comparison of the three numerical results, in order to establish the formalism as a numerically robust tool. This allow to envisage applications using state--of--the--art structure inputs, such as  the Dispersive Optical Model (DOM)~\cite{Dickhoff17}. 

The outline of the paper is as follows. In Section \ref{Reaction_Theory}, we present a description of the form	alism used in this work (\ref{Reaction}) and also provide a more detailed derivation of the surrogate cross section (\ref{surrogates}) as well as of the mechanism for compound system decay (\ref{decay}). In Section \ref{Structure} we present how nuclear structure considerations enter the formalism through the use of a dispersive optical potential, and in Section \ref{results} the results of this work are presented. A summary and further discussion is provided in Section \ref{Discussion}.

%% file: Formalism.tex
Let us consider the reaction $A(d,p)B^*$ which includes elastic breakup and any other inelastic processes. Several recent works \cite{Lei:15,Potel:15b,Lei:15b,Carlson:15} have reviewed the formalism in detail, thus here we briefly compiled the equations necessary for our subsequent applications and discussion.

As in \cite{Ichimura:85,Austern:87}, we will adopt a spectator approximation for the proton, which means the proton--target interaction will not explicitly excite the target $A$. We will thus  substitute the proton--target interaction $V_{Ap}(r_{Ap},\xi_A)$ by an optical potential $U_{Ap}(r_{Ap})$. 
The Hamiltonian for the problem is
\begin{align}\label{eqForm1}
\nonumber H&=K_n+K_p+h_A(\xi_A)+V_{pn}(r_{pn})\\
&+V_{An}(r_{An},\xi_A)+U_{Ap}(r_{Ap}),
\end{align}
where $K_n$ and $K_p$ are the kinetic energy operators acting on the neutron and  proton coordinates respectively.  Defining the total wavefunction $\Psi^{(+)}(\mathbf r_p,\mathbf r_n,\xi_A)$ with outgoing boundary conditions at a given energy $E$ satisfying
\begin{eqnarray}\label{eqForm2}
\left(H-E\right)\Psi^{(+)}(\mathbf r_p,\mathbf r_n,\xi_A)=0,
\end{eqnarray} 
and the eigenfunctions $\varphi_B^c(\mathbf r_n,\xi_A)$ of the $B$ nucleus, we can write the inclusive cross section in the post representation as an explicit sum over all energy--allowed states $c$,

\begin{align}\label{eqForm3}
\nonumber  \frac{d^2\sigma}{d\Omega_pdE_p}&=\frac{2\pi}{\hbar v_d}\rho_p(E_p)\\
\nonumber&\times\sum_c\left|\left\langle\chi_f^{(-)}\varphi_B^c|V_{pn}+V_{pA}-U_{pB} \,|\Psi^{(+)}\right\rangle\right|^2\\
&\times\delta(E{-}E_p{-}\varepsilon_B^c),
\end{align}
where $\chi_f^{(-)}(\mathbf r_p)$ is the scattering wave of the emitted proton, solution  of the Scrh\"odinger equation with the distorting potential $U_{Ap}(r_{Ap})$, and $\varepsilon_B^c$ is the energy of the state $\varphi_B^c$. The proton level density is given by  
\begin{eqnarray}\label{eqForm4}
\rho_p(E_p)=\frac{m_p k_p}{8\pi^3\hbar^2},
\end{eqnarray}
 where $E_p$ and $k_p$ are the kinetic energy and momentum of the detected proton, respectively. The neutron--target wavefunction $\varphi_B^c(\mathbf r_n,\xi_A)$ can be obtained by projecting the solution of a Lippmann--Schwinger equation in the breakup channel over the proton coordinates,
 \begin{eqnarray}\label{eqForm5}
\varphi_B(\mathbf r_n,\xi_A)=\left[E-K_p-H_B\right]^{-1}\left(\chi_f^{(-)}\right|V_{pn}\,\left|\,\vphantom{\chi_f^{(-)}}\Psi^{(+)}\right\rangle,
 \end{eqnarray}
where the curved bracket denotes integration over the proton coordinates only. It can be shown that, as a consequence of flux conservation, the total amount of breakup can be divided into an elastic (EB) and non--elastic (NEB) contribution,
    \begin{eqnarray}\label{eqForm7}
    \int\nabla\left(\varphi_B\nabla\varphi_B^{*}-\varphi_B^{*}\nabla\varphi_B\right)d\mathbf{r}_{n}-2i\int|\varphi_B|^2  W_{An}d\mathbf{r}_{n} \nonumber \\
    = 2i\Im\int \varphi_B \left(\chi_f^{(-)}\right|V_{pn}\,\left|\,\vphantom{\chi_f^{(-)}}\Psi^{(+)}\right\rangle^\dagger d\mathbf{r}_{n},
    \end{eqnarray}
  where $ W_{An}$  is the imaginary part of the neutron--target optical potential.
 The first term in the l.h.s. can be cast into an outgoing elastic flux across the surface enclosing the volume, while the second term accounts for the non--elastic breakup in the post, Ichimura--Austern--Vincent (IAV) formalism. The double differential cross section for that non-elastic contribution can then be written as,
 \begin{eqnarray}\label{eqForm6}
 \left.\frac{d^2\sigma}{d\Omega_pdE_p}\right]^{IAV}=-\frac{2}{\hbar v_d}\rho_p(E_p)\left\langle\vphantom{\chi_f}\varphi_B\right|  W_{An}\, \left|\vphantom{\chi_f}\varphi_B\right\rangle.
 \end{eqnarray}
 The last expression gives the exact NEB within the spectator approximation implicit in the Hamiltonian (\ref{eqForm1}), to the extent that the wavefunction $\Psi^{(+)}$ used in (\ref{eqForm5}) is the exact wavefunction, as expressed by (\ref{eqForm2}). We can approximate instead the exact result with the wavefunction in the asymptotic entrance channel,
   \begin{eqnarray}\label{eqForm10}
\Psi^{(+)}(\mathbf r_p,\mathbf r_n,\xi_A)\approx\chi_i(\mathbf r_d)\,\phi_d(r_{pn})\phi_A(\xi_A),
   \end{eqnarray}
 where $\chi_i(\mathbf r_d)$ is the distorted wave of the incoming deuteron (defined by an optical potential $U_{Ad}(r_{Ad})$), and $\phi_d(r_{pn})$,  $\phi_A(\xi_A)$   are the intrinsic wavefunctions of the deuteron, and of nucleus $A$, respectively. 
 
 One can also derive an equivalent expression in the prior (Udagawa--Tamura, UT) formalism \cite{Udagawa:81}. Defining
  \begin{eqnarray}\label{eqForm8}
  \varphi_B^{UT}(\mathbf r_n)=\left[E-K_p-H_B\right]^{-1}\left(\chi_f^{(-)}\right|V_{prior}\,\left|\,\vphantom{\chi_f^{(-)}}\chi_i\,\phi_d\right\rangle,
  \end{eqnarray}
with  
   \begin{eqnarray}\label{eqForm9}
V_{prior}=U_{An}(r_{An})+U_{Ap}(r_{Ap})-U_{Ad}(r_{Ad}),
   \end{eqnarray} 
   and the Hussein--McVoy (HM) term,
      \begin{eqnarray}\label{eqForm11}
   \psi_B^{HM}=\left(\chi_f^{(-)}\right|\,\left.\,\vphantom{\chi_f}\phi_d\,\chi_i\vphantom{\chi_f^{(-)}}\right\rangle,
      \end{eqnarray} 
it can be proved that

\begin{eqnarray}\label{eq-all}
\nonumber & \left.\frac{d^2\sigma}{d\Omega_pdE_p}\right]^{IAV}&=\left.\frac{d^2\sigma}{d\Omega_pdE_p}\right]^{UT}+\left.\frac{d^2\sigma}{d\Omega_pdE_p}\right]^{HM}
\end{eqnarray}
where the first term reflects explicitly the 2-step process of elastic breakup followed by capture, and corresponds to  the prior formulation of UT \cite{Udagawa:81}:
\begin{eqnarray}\label{eq-UT}
 & \left.\frac{d^2\sigma}{d\Omega_pdE_p}\right]^{UT}&=-\frac{2}{\hbar v_d}\rho_p(E_p)\Im\left\langle\,\psi_B^{UT}|W_{An}\,|\psi_B^{UT}\right\rangle .
\end{eqnarray}
 The second is related to the HM formulation and can be attributed to non--orthogonality corrections:
\begin{align}\label{eq-NO}
\nonumber  \left.\frac{d^2\sigma}{d\Omega_pdE_p}\right]^{NO}&=\Re\left\langle\,\psi_B^{{HM}}|W_{An}|\psi_B^{UT}\right\rangle\\ &+\left.\left\langle\,\psi_B^{{HM}}|W_{An}\,|\psi_B^{{HM}}\right\rangle\right].
\end{align}

Naively, one may think that the NO term is small and can be neglected, making Eq.(\ref{eqForm6}) and Eq.(\ref{eq-UT}) identical.  However, we show in Sect. \ref{results} that the non-orthogonality contribution is not negligible, and indeed contains subtle interferences. It can even become the dominant term. 

For some applications, as well as from a theoretical point of view, it is interesting to disentangle the different contributions to the NEB cross section. When using $(d,p\gamma)$ reactions as surrogates for neutron capture, for example, one would need to assess the fraction of NEB that corresponds to the formation of a compound neutron--target system. Within this context, some authors have proposed to define a fusion radius $r_F$, such that absorption from the interior part of the optical potential leads to fusion, while peripheral absorption gives rise to other inelastic processes \cite{Udagawa:85}. This approach has been used recently in the context of surrogate reactions, \cite{Ducasse:16} but it usually requires to previously fit $r_F$ from experimental $(n,\gamma)$ data. A more sophisticated approach is to calculate explicitly the portion of the Hamiltonian responsible for direct and pre--equilibrium reactions, within, e.g., RPA or QRPA (see, e.g., \cite{Dupuis:11,Blanchon:15}). This contribution could then be substracted from the total absorption in order to single out the compound nucleus formation. 

%% file: Application_Surrogates.tex
Neutron capture reactions of interest to astrophysics are typically dominated by the compound--reaction mechanism. 
The appropriate formalism for the description of a compound--nuclear (CN) capture reaction $n+A \rightarrow B^*$ $\rightarrow \gamma+B$ 
%(for example, $n$ + $^{48}Ca$ $\rightarrow$ \yeightx$^*$ $\rightarrow$ $^{49}$Ca + $\gamma$ - find better example)
is a statistical one 
\cite{HauserFeshbach:52}
. When a large number of levels are populated in the compound system, the energy--averaged cross sections can be calculated in the framework of the Hauser--Feshbach formalism, which properly takes account of the conservation of angular momentum and parity in the reaction (width fluctuations omitted here): 
\begin{align}
\sigma_{n+A, \chi}(E_{n})&= \sum_{J,\pi}  \sigma_{n+A}^{CN}(E_{ex},J,\pi) \\ &\times G_{\chi}^{CN}(E_{ex},J,\pi) \; ,
\label{eq:DesReact}
\end{align}
\noindent
where $\chi$ denotes the exit channel of interest, here $\chi=\gamma+B$, and
the excitation energy $E_{ex}$ of the compound nucleus, $B^*$, is related to the center-of-mass energy $E_n$ via the neutron separation energy $S_n(B)$. 
% $S_n$: $E_n=E-S_n(B)$. 
Near stability, the formation cross section $\sigma_{n+A}^{CN} = \sigma (n+A \rightarrow B^*)$ can be calculated to a reasonable accuracy by using existing optical potentials, while the theoretical decay probabilities $G_{\chi}^{CN}$ for the different decay channels $\chi$ are often quite uncertain. The latter are difficult to calculate accurately since they require knowledge of level densities, and strength functions for the various possible exit channels \cite{Capote:09}.  The objective of the surrogate method is to determine or constrain these decay probabilities experimentally.

In a surrogate (d,p) experiment, the compound nucleus $B^*$ is produced via the $d+A \rightarrow p+B^*$ reaction, and the desired decay channel $\chi$ is observed in coincidence with the outgoing proton $p$.  
The probability for forming $B^*$ in the surrogate reaction (with specific values for $E_{ex}$, $J$, $\pi$) is $F_{\delta}^{CN}(E_{ex},J,\pi,\theta_b)$, where $\delta$ refers to the entrance channel reaction $A(d,p)$ and $\theta_b$ is the angle of the outgoing proton $p$ relative to the beam axis. The quantity
\begin{equation}
P_{\delta\chi}(E_{ex},\theta_b) = \sum_{J,\pi} F_{\delta}^{CN}(E_{ex},J,\pi,\theta_p) \;\; G_{\chi}^{CN}(E_{ex},J,\pi) \; ,
\label{Eq:SurReact}
\end{equation}
which gives the probability that the compound nucleus $B^*$ was formed with energy $E_{ex}$
and decayed into channel $\chi$, can be obtained experimentally, by measuring $N_{\delta}$, the total number of surrogate events, and $N_{\delta\chi}$, the number of coincidences between the direct-reaction particle and the observable that identifies the relevant exit channel:
$P^{exp}_{\delta\chi}(E_{ex},\theta_p) = N_{\delta\chi}(E_{ex},\theta_p) / N_{\delta}(E_{ex},\theta_p)\epsilon_{\delta} (E_{ex})\; .$
Here, $\epsilon_{\delta}(E_{ex})$ denotes the efficiency for detecting the exit-channel $\chi$ (when in coincidence with the outgoing proton $p$).

The distribution $F_{\delta}^{CN}(E_{ex},J,\pi,\theta_b)$, which may be very different from the CN spin-parity populations following the absorption of the neutron in the desired reaction
%~\footnote{
%This is often referred to as the `spin-parity mismatch' between the desired and surrogate reactions.
%}
, has to be determined theoretically, so that the branching ratios $G_{\chi}^{CN}(E_{ex},J,\pi)$ can be extracted from the measurements \cite{Forssen:07,EscherDietrich:10,Scielzo:10a}.  In practice, the decay of the CN is modeled and the $G_{\chi}^{CN}(E_{ex},J,\pi)$ are obtained by adjusting parameters in the model to reproduce the measured probabilities $P_{\delta\chi}(E_{ex},\theta_b)$ \cite{Escher:12rmp}. Subsequently, the sought-after cross section can be obtained by combining the calculated cross section $\sigma_{\alpha}^{CN}(E_{ex},J,\pi)$ for the formation of $B^*$ (from $n+A$) with the extracted decay probabilities $G_{\chi}^{CN}(E_{ex},J,\pi)$, see Eq. \ref{eq:DesReact}.

%Some points to mention here: formalism applicable also to $(p,\gamma)$, (n,p), etc., comments about the challenges when moving off stability: optical potentials, direct capture contributions.

%\section*{Acknowledgments}
%This work was performed in part under the auspices of the U.S. Department of Energy by Lawrence Livermore National Laboratory under contract DE-AC52-07NA27344 (LDRD 16-ERD-022 and ASC-PEM projects).

%% file: Compound_decay.tex
%Contributed by J. Escher
To calculate compound-nuclear reaction cross sections, one needs to have information on discrete levels and level densities in the residual nuclei, $\gamma$-ray transmission coefficients (strength functions), and particle (e.g. neutron, proton) transmission coefficients~\cite{Capote:09}.  Where fission plays a role, a description of fission barriers and associated level densities is required as well.  
%In recent years, much progress has been made to predict many of these inputs using microscopic theories.  To achieve reliable reaction predictions, it is important to complement these predictions with experimental data. Measurements of neutron resonance spacings, for instance, provide stringent constraints on level densities and $\gamma$-induced reactions provide useful information on $\gamma$-ray strength functions.  In addition, indirect methods, such as the surrogate \textcolor{red}{[reference??]} and Oslo \textcolor{red}{[reference???]} methods, aim at extracting relevant information from observing decays of compound nuclei produced by alternative reactions (e.g. from inelastic scattering, transfer reactions, or $\beta$-decay)~\cite{Escher:16a}. 
%In turn, when the decay of a compound nucleus is well-characterized, it is possible to gain insights into the processes that formed the CN.  For example, the decay of a given compound nucleus depends on its angular-momentum and parity distribution prior to decay, as will be discussed for the case of $^{41}$Ca below.

The decay probability $G_{\chi}^{CN}(E_{ex},J,\pi)$ (Eq.~(\ref{eq:DesReact})) is the branching ratio for the decay of the compound state into the desired exit channel $\chi$,
integrated over a suitable energy range in the residual nuclei 
(for the example of capture, the integration is taken over the energy spectrum of primary $\gamma$ rays emitted from the compound nucleus).  
It contains transmission coefficients for the competing exit channels as well as the associated level densities, and information on discrete levels:
\begin{strip}
\begin{eqnarray}  
	\frac{dG_{\chi}^{CN}(E_{ex},J,\pi)}{dE_{\chi}} = \sum_{\ell' s' I'} \frac{T^J_{\chi \ell' s'} \rho_{I'}(U') }
{\sum'_{\chi'' \ell'' s''} T^J_{\chi'' \ell'' s''}  + \sum_{\chi'' \ell'' s'' I''} \int T^J_{\chi'' \ell'' s''} (E_{\chi''}) \rho_{I''} (U'') dE_{\chi''}}.
\label{Eq:HFGeneral}
\end{eqnarray}
\end{strip}
The quantities $\ell'$ and $\ell''$ are the relative orbital angular momentum in the exit channels.
The transmission coefficients are written as $T^J_{\alpha l s}$ and $\rho_{I'}(U')$ denotes the density of levels of spin $I'$ at excitation energy $U'$ in the residual nucleus.  All energetically possible final channels $\chi''$ have to be taken into account, thus the denominator includes contributions from decays to discrete levels in the residual nuclei (given by the first sum in the denominator, $\sum'$) as well as contributions from decays to regions described by a level density in the residual nuclei (given by the second sum in the denominator which involves an energy integral of transmission coefficients and level densities in the residual nuclei).  
%Width fluctuation corrections $W_{\alpha \chi}$ are included in order to account for correlations between the incident and outgoing reaction channels~\cite{Moldauer:61,Hilaire:03}.
In writing Eq.~(\ref{Eq:HFGeneral}), we have suppressed the parity quantum number except for that of the compound nucleus.  In fact, the level density depends in principle on parity and all sums over quantum numbers must respect parity conservation.
% Contributed by G. Perdikakis

In the particular scenario in which the compound nucleus de-excites via emission of $\gamma$ rays after it has been populated via a $(d,p)$ reaction, the formation -as discussed earlier- of the compound system through a deuteron-induced reaction may lead to a generally different spin distribution than in the case of a neutron- or proton-capture reaction. The decay of the compound system once formed, however, can be described for any given spin using the same approach as in any neutron (or proton) capture reaction. The $\gamma$ decay probability $G_{\gamma}^{CN}(E_{ex},J,\pi)$ is connected in the Hauser-Feshbach picture to statistical quantities of the decaying compound system, namely the density of levels, and the $\gamma$ decay strength distributions. The $\gamma$ decay strength function for $\gamma$ rays of a given multipolarity is related to the average reduced partial $ \gamma $--radiation width for $\gamma$ decay $ \cev{\Gamma}^J_{XL}(\epsilon_\gamma) $ through the equation
\begin{eqnarray}
f^J(\epsilon_\gamma)=\frac{\cev{\Gamma}^J_{XL}(\epsilon_\gamma)}{\epsilon_\gamma^{2L+1}D_l}
\label{eq:DesGammaDecay}
\end{eqnarray}
where $ L $ is the multipolarity of the $\gamma$ ray, $ X $ is the type of multipole ($ E $ for Electric, $ M $ for magnetic), and $ D_l $ is the average level spacing in the approximate excitation energy neighborhood of the nucleus before the $\gamma$ emission (in other words 1/$ D_l $ is equal to the average level density in the compound system before emission). The factor $ \epsilon_\gamma^{2L+1} $ provides the known energy dependence as a function of multipolarity.

%The decay of the compound nuclear system can be calculated in the Hauser-Feshbach formalism through the decay probabilities  $G_{\chi}^{CN}(E_{ex},J,\pi)$ (see Eq.\ref{eq:DesReact}). In general,the exit channel of interest, $ \chi $, can be a nucleon or gamma decay channel. Here, however, we will concentrate 

%% file: Optical_potentials_related_quantities.tex
%One of the underlying science questions of rare isotope physics is how the properties of protons and neutrons 
%in the nucleus change from the valley of stability to the respective drip lines.
%Elastic nucleon scattering has traditionally provided insights into this question for stable nuclei by clarifying how a 
%nucleon experiences its propagation through the nucleus at positive energy.
%This experience is usually represented in terms of an energy-dependent complex potential, the optical potential.
%A theoretical framework for this potential was developed by Feshbach~\cite{Feshbach:58,Feshbach:62} emplying projection techniques.
%A related connection between elastic-scattering data and nucleon propagation was established by Bell and Squires~\cite{Bell:59} demonstrating that the nucleon elastic-scattering $\mathcal{T}$-matrix is equivalent to the reducible self-energy obtained by iterating the irreducible one to all orders with the free nucleon propagator~\cite{Villars67, BlaizotR86,Dickhoff08}.
%This provides a more flexible approach in the present context since both reaction and structure information is simultaneously addressed unlike the Feshbach projection formulation which only emphasizes the elastic scattering aspects of the optical potential~\cite{Feshbach:58,Feshbach:62}.

By simultaneously studying the propagation of a nucleon through the nucleus at positive energy generating experimentally accessible elastic scattering cross sections, as well as the movement of nucleons in the bound states at negative energy, it is possible to shed light on the fundamental question of how the properties of protons and neutrons in the nucleus change
from the valley of stability to the respective drip lines.
Detailed knowledge of this propagation process at positive energies allows for an improved 
description of other hadronic reactions, including those that purport to 
extract structure information, like transfer or knockout reactions.
Structure information associated with the removal of nucleons
from the target nucleus is therefore a subject of these 
studies and must be supplemented by the appropriate description of the 
hadronic reaction utilized to extract it.

%Traditionally, positive energy nucleons are described by fitted optical potentials mostly in local form~\cite{Varner91,Koning:03}.
%Bound nucleons are usually analyzed with static potentials that lead to an independent-particle model (IPM) modified by the interaction between valence nucleons as \textit{e.g.} in traditional shell-model calculations~\cite{Brown01,Caurier05}.
%The link between nuclear reactions and nuclear structure is provided by considering these potentials as representing  different energy domains of one underlying nucleon self-energy.
%The seminal work of Mahaux and Sartor emphasized the link between these traditionally separate fields in nuclear physics~\cite{Mahaux86,Mahaux:91}.
The main idea behind the  dispersive optical model (DOM) approach is to employ the concepts of the 
Green's function formulation of the many-body problem~\cite{Dickhoff08} to allow experimental
data to constrain the static and dynamic content of the nucleon self-energy through the use of dispersion relations.
By employing dispersion relations, the method provides a critical link between the physics above and below the Fermi energy with both sides being influenced by the absorptive potentials on the other side.
Since the self-energy determines both the properties of the system when 
a nucleon is removed as well as when one is added to the ground state (of a 
target), a unique link between structure and (initially elastic) scattering 
information can be forged.

The St. Louis group made an initial foray into this approach by extending the DOM application to a simultaneous analysis of different nuclei belonging to an isotope chain like the calcium isotopes~\cite{Charity06,Charity07,Mueller11}.
Such an approach is therefore ideally suited to study rare isotopes by providing data-constrained extrapolations into unknown territory which can subsequently be probed by new experiments in inverse kinematics.
These initial developments still employed local potentials and the results of Ref.~\cite{Mueller11} for ${}^{40}$Ca and ${}^{48}$Ca have been extrapolated to ${}^{60}$Ca and are employed in the present work. 

Further insights into this approach are provided by \textit{ab initio} Green's function calculations or other many-body techniques that clarify the appropriate functionals that are needed to describe the essential features of the nucleon self-energy~\cite{Waldecker2011,Dussan11}.
Recently a new step has been taken motivated by these \textit{ab initio} calculations by introducing fully non-local absorptive potentials in the analysis of ${}^{40}$Ca~\cite{Mahzoon14}.
This work is currently being extended to ${}^{48}$Ca, ${}^{208}$Pb and other systems to provide important ingredients for the analysis of
nuclear reactions both on and off stability \cite{Dickhoff17}.   

%% file: results.tex
\subsection{Benchmark and numerical details}
In order to benchmark the three implementations of the breakup--fusion reaction formalism, we have chosen the $^{93}$Nb($d,p$) reaction, for which agreement with  experimental data has been verified (\cite{Pampus:78,Kleinfeller:81,Mastroleo:90,Lei:15,Potel:15b}). We have used  a deuteron global optical model potential \cite{Han:06} to describe the Nb--deuteron interaction $U_{Ad}(r_{Ad})$, and the Koning--Delaroche global nucleon--nucleus optical potential \cite{Koning:03} to describe the proton--Nb ($U_{Ap}(r_{Ap})$) and neutron--Nb ($U_{An}(r_{An})$) interactions. The deuteron ground-state wave function is taken to be
an $\ell = 0$ state with a radial wave function generated by
a Woods--Saxon potential with radius $R_d = 0.4$ fm and
diffuseness $a_d$ = 0.6 fm. When the real depth is adjusted to reproduce the binding energy of the deuteron, the resulting
wave function is compatible with the experimental value of
the mean--square radius of the deuteron and the zero--range
constant $D_0 =-122.5$ MeV fm$^{3/2}$.

\begin{figure*}[h]
\centerline {
\includegraphics*[width=13cm]{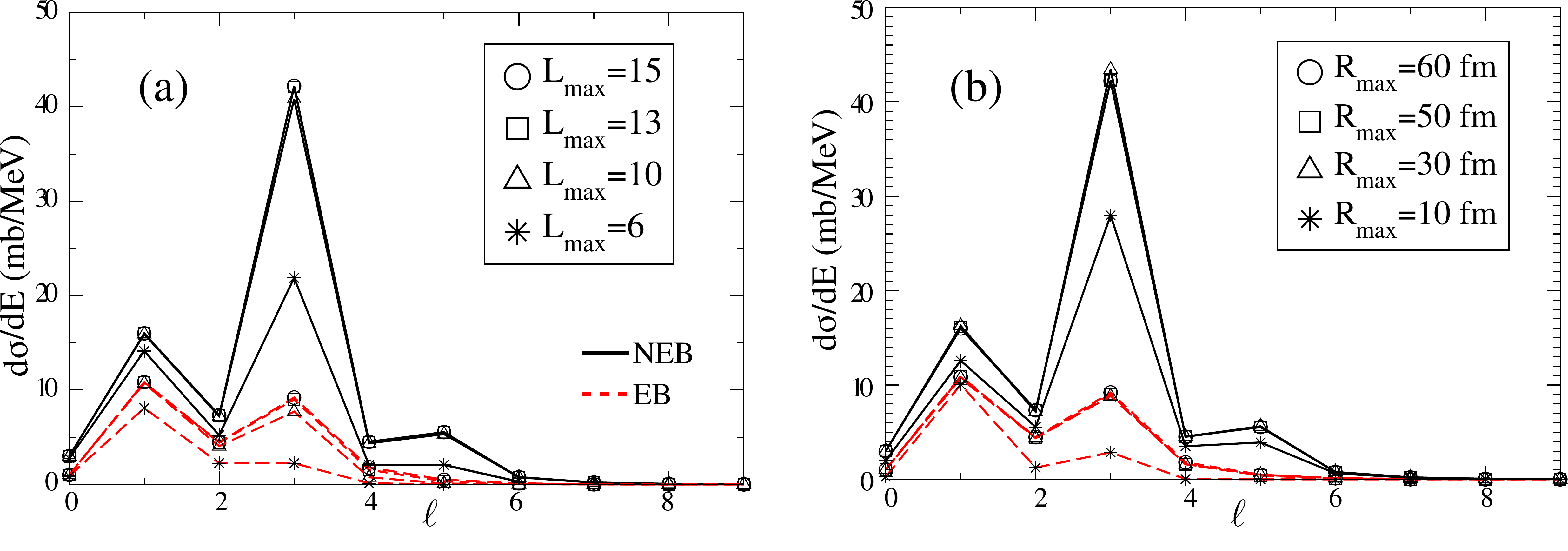}
}
\caption{Transferred angular momentum in the $^{93}$Nb$(d,p)$ reaction with a beam of $E_d=10$ MeV deuterons. In the panel (a) we illustrate the convergence of the NEB (solid black line) and EB (dashed red line) calculation with respect to the number of partial waves $L_{max}$, while in (b) we show the dependence of the same results on the upper limit $R_{max}$ of radial integrations. Panel (a) have been calculated using $R_{max}=60$ fm, and panel (b) with $L_{max}=15$.}
\label{convergence}
\end{figure*}

The numerical implementation relies on a partial wave decomposition of the scattering channels involved (i.e., the incoming deuteron and outgoing proton channels), and the number $L_{max}$ of partial waves is increased until convergence is reached.   The upper limit $R_{max}$ of the radial integrations is also tested for convergence, and one finds that it depends on whether the calculation is made in the UT or IAV formalism.   It is generally found that numerical convergence with respect to $L_{max}$ and $R_{max}$ of the EB contribution is slower than for the NEB one. In Fig. \ref{convergence} we illustrate the convergence with respect to $L_{max}$ and $R_{max}$ for the $^{93}$Nb$(d,p)$ reaction with a beam of $E_d=10$ MeV deuterons, performed in the UT formalism.

\begin{figure*}[h]
\centerline {
\includegraphics*[width=13cm]{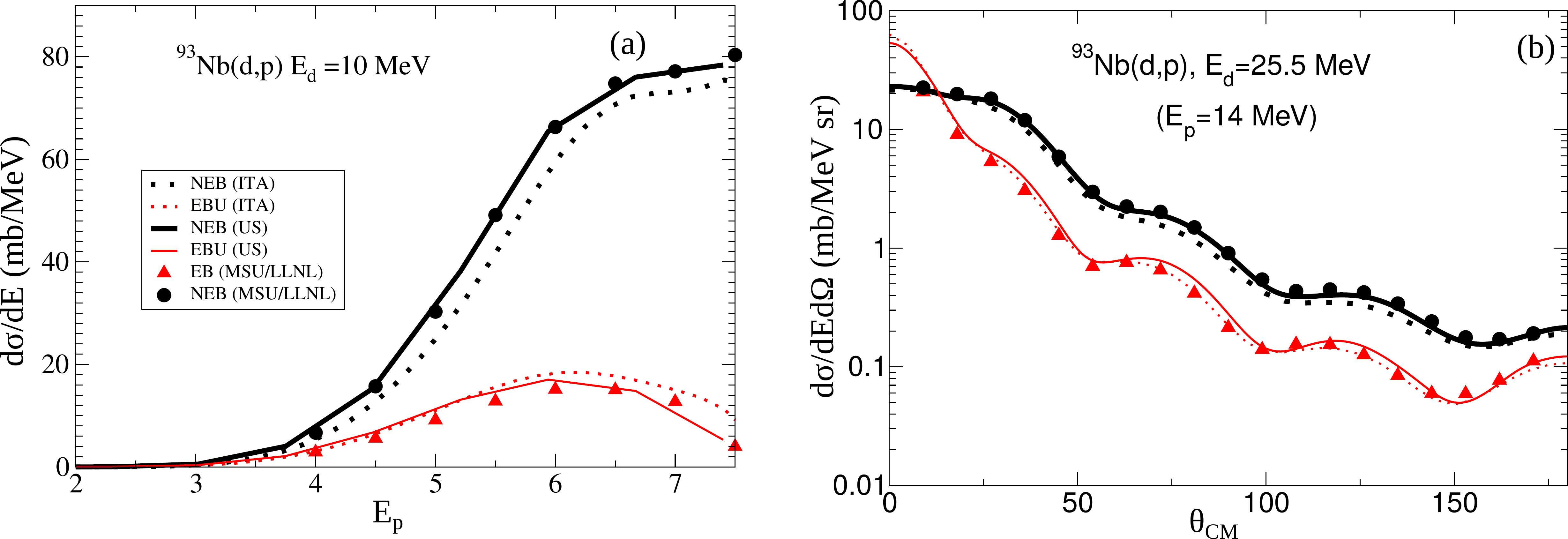}
}
\caption{Benchmark calculations comparing the three codes mentioned in the text. We find a very good agreement between US and MSU/LLNL for NEB (thick black line and black dots, respectively). The zero--range approximation (ITA, thick black dotted line) is found to be a very good one in the cases considered here. Good agreement is also found between the EB calculations performed with US (thin red line), ITA (thin red dotted line) and MSU/LLNL (red triangles).}
\label{fig_bench}
\end{figure*}

\begin{figure*}[h]
\centerline {
\includegraphics*[width=13cm]{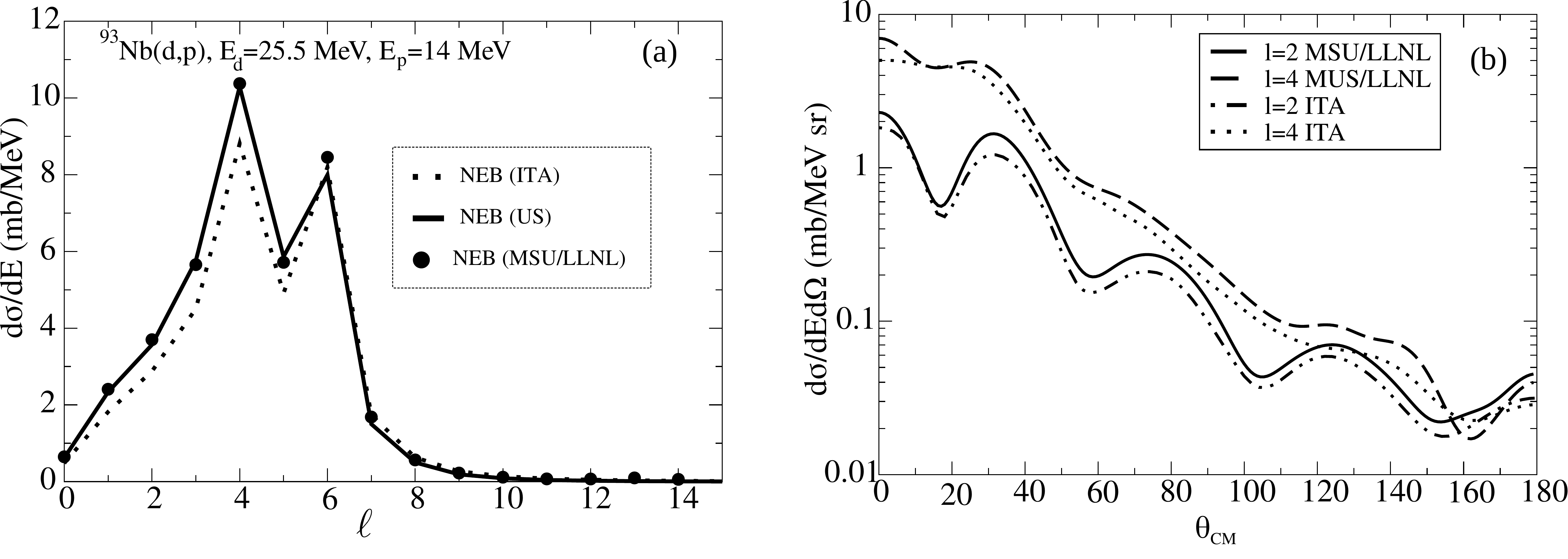}
}
\caption{NEB cross sections for the different neutron angular momenta. In panel (a) we benchmark the three codes against the angle--integrated differential cross section for the neutron angular momenta contributing to the total cross section. In panel (b) we compare the angular shapes of two selected neutron angular momenta, as predicted by the zero range (ITA) and finite range (MSU/LLNL) calculations.}
\label{fig_bench2}
\end{figure*}

In Fig. \ref{fig_bench} we show  benchmark calculations using the three codes, regarding differential ($d\sigma/dE$, panel (a)) and double differential ($d^2\sigma/dEd\Omega$, panel (b)) cross sections at two different beam energies ($E_d=10$ MeV in panel (a), $E_d=25.5$ MeV in panel (b)), for the reaction $^{93}$Nb$(d,p)$. In Fig.~\ref{fig_bench2}, we show more detailed results of the comparison between the calculations of the $^{93}$Nb$(d, p)$ NEB cross section at 25 MeV, which in all cases correspond to emission of a proton of 14 MeV. In Fig. 2a, we display the benchmark calculations of the partial-wave decomposition of the differential NEB cross section. The structure in the cross section roughly corresponds to the structure in the neutron absorption cross section at the same energy. The two finite-range calculations agree quite well but are about 10\% larger than the zero-range calculation for low values of the orbital angular momentum. This difference also appears in Fig. 2b, where we see not only differences in magnitude but also slight differences in the angular dependence of the zero-range and finite range calculations of the NEB proton angular distributions corresponding to the $\ell=2$ and $\ell=4$ partial waves of Fig. 2a. We note that the NEB cross sections are strongly localized in angle, falling by an order of magnitude in 60$^\circ$. We also emphasize that the proton angular distribution is not the same for each partial wave. For the two cases shown, the $\ell=2$ partial wave shows a diffractive structure that reflects absorption from the internal region, while the more superficial $\ell=4$ partial wave angular distribution displays a smoother decrease with angle.

Zero--range calculations were also performed by the University of Seville group, using the same method used to perform the finite-range calculations, and show much better agreement with the finite--range ones than those of the ITA group~\cite{Lei:15}. Further testing will be necessary to pinpoint the differences in the two methods used that give rise to the differences in the results. In general, we obtain a good agreement of the calculations despite the differences of implementation. 
\subsection{Application to the Ca isotopic chain}
The methods benchmarked in the previous section are now applied to $(d,p)$ reactions on Ca isotopes. To illustrate the systematics as one moves away from stability we take the double magic stable $^{40}$Ca isotope, and counterpose it to results with $^{48}$Ca and with $^{60}$Ca, predicted to be at the neutron dripline \cite{Hagen:12}. The beam energy  ranges from $E_d=10$ to 40 MeV since we expect these energies will be available experimentally at the new Facility for Rare Isotope Beams and offer the optimum window for large cross sections in transfer. Unless otherwise stated, results are obtained with the DOM potential discussed in Section \ref{Structure}. 
The deuteron state is identical to the one described  in the previous section. We use $L_{max}=15$ and $R_{max}=60$ fm, for which good convergence of the results has been verified. 
\begin{figure*}[h]
\centerline {
\includegraphics*[width=16cm]{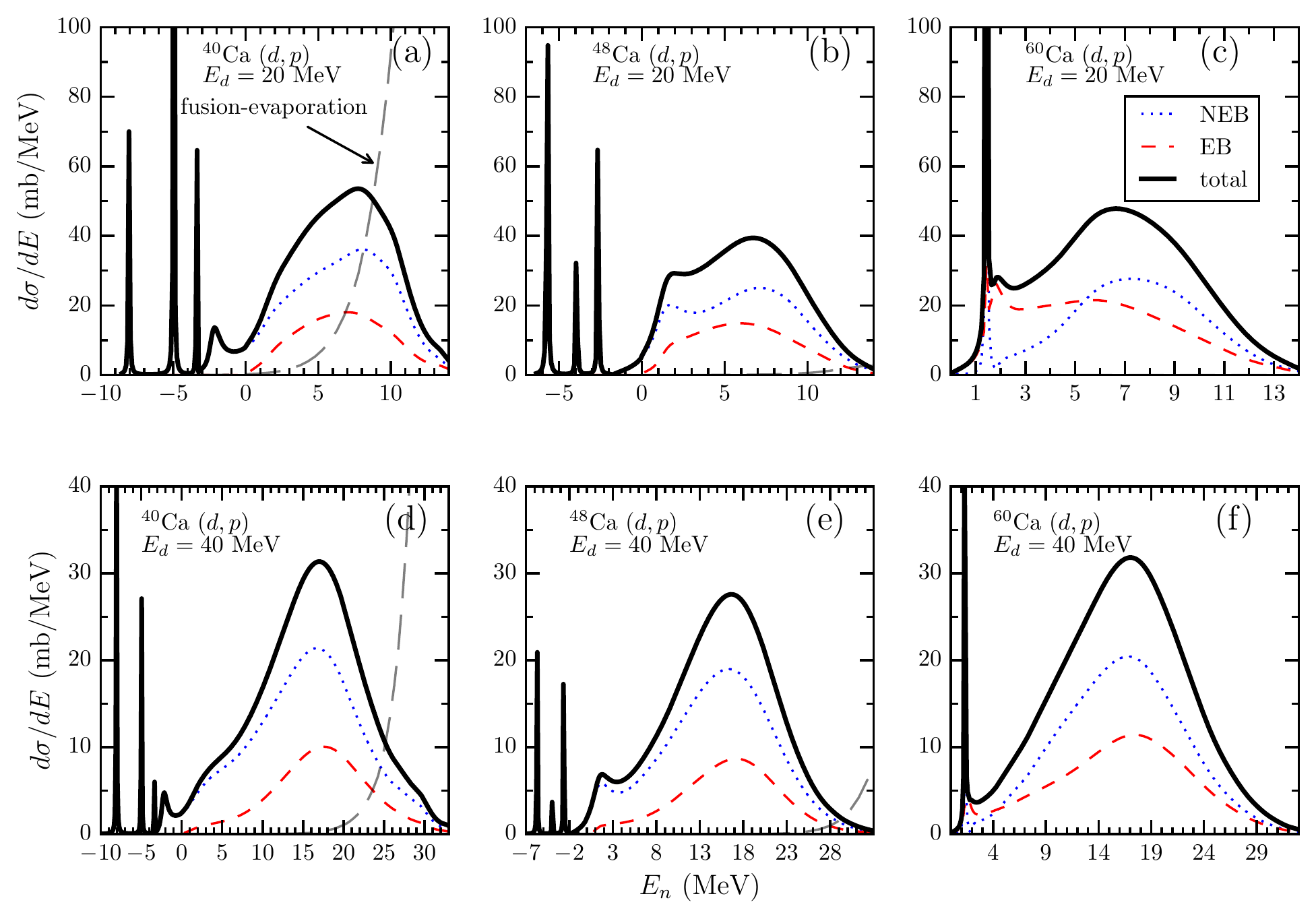}
}
\caption{Elastic breakup (EB) and non--elastic breakup (NEB) proton spectra for the reactions $^{40}$Ca$(d,p)$, $^{48}$Ca$(d,p)$, and $^{60}$Ca$(d,p)$, for beam energies $E_d=20$ MeV and $E_d=40$ MeV. The proton singles cross sections are measured in mb/MeV, as a function of the transferred neutron energy $E_n$. For comparison, we show with a long--dashed line the contribution from the fusion--evaporation mechanism to the proton spectra, as calculated with a Hauser-Feshbach code.}
\label{fig_ebneb}
\end{figure*}

In Fig. \ref{fig_ebneb} we show the neutron energy distributions for $(d,p)$ reactions at $E_d=20$ MeV and $40$ MeV. We compare the elastic breakup component (red dashed line) to the non--elastic component (blue dotted line) and the total cross section (solid line). Panels (a) and (d) refer to $^{40}$Ca, panels (b) and (e) refer to $^{48}$Ca and panels (c) and (f) refer to $^{60}$Ca. We note that for all cases studied both EB and NEB have significant contributions. As was seen for $^{93}$Nb,  the EB component for $^{40}$Ca is about one third of the total strength, with NEB contributing with roughly 2/3. This is also the case for $^{48}$Ca. Whereas  for the stable isotopes the non--elastic component is always dominant, as one reaches the limits of stability the elastic breakup becomes more important, particularly close to threshold. It has to be noted that a contribution of low energy protons evaporated after fusion of the deuteron with the target may be present in the high energy part of the spectrum. This mechanism can be estimated independently within the Hauser--Feshbach formalism, and it is shown in the figure with the long--dashed line. The contribution is sizeable for the $ ^{40} $Ca case, but as we move away from stability and the $Q$--value for proton evaporation becomes smaller, the contribution diminishes, becoming vanishingly small for $^{60}$Ca. 
%To better illustrate this effect, percentage contributions of EB relative to the total cross section are presented in Fig XXX. 
\begin{figure*}[h]
	\centerline {
		\includegraphics*[width=16cm]{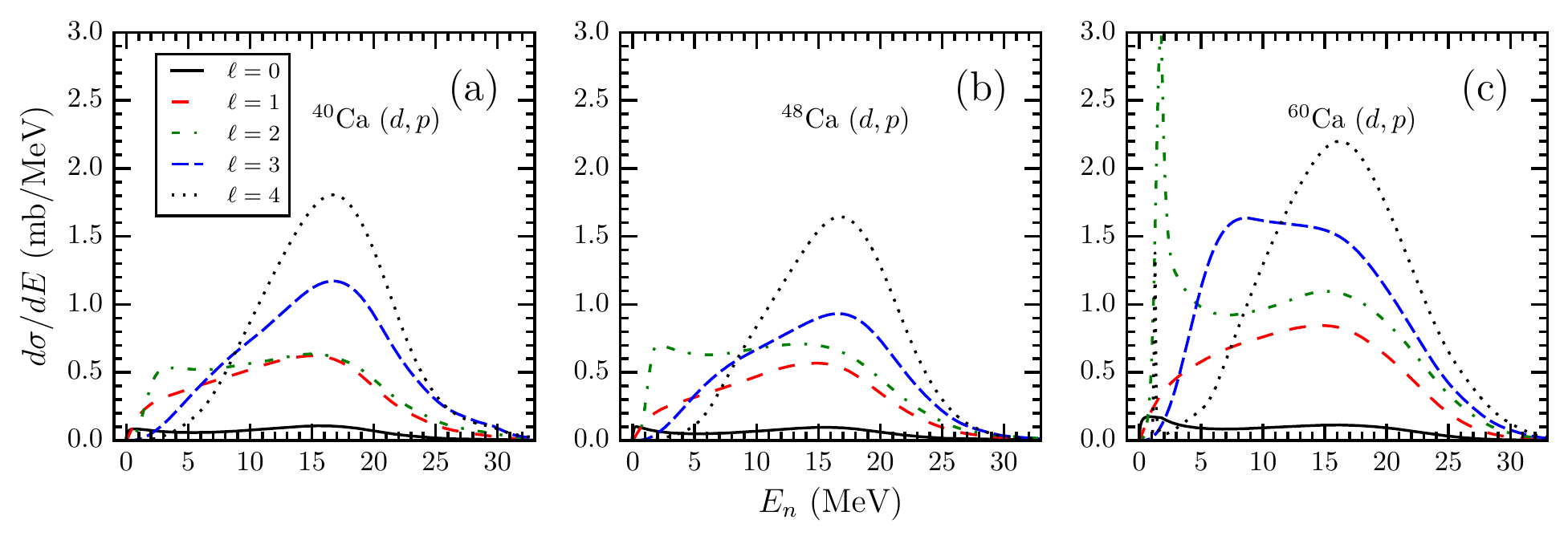}
	}
	\caption{Population of different final neutron angular momentum states in EB processes for the reactions $^{40}$Ca$(d,p)$, $^{48}$Ca$(d,p)$, and $^{60}$Ca$(d,p)$. The beam energy is $E_d=40$ MeV.}
	\label{fig_ebl}
\end{figure*}
\begin{figure*}[h]
\centerline {
\includegraphics*[width=16cm]{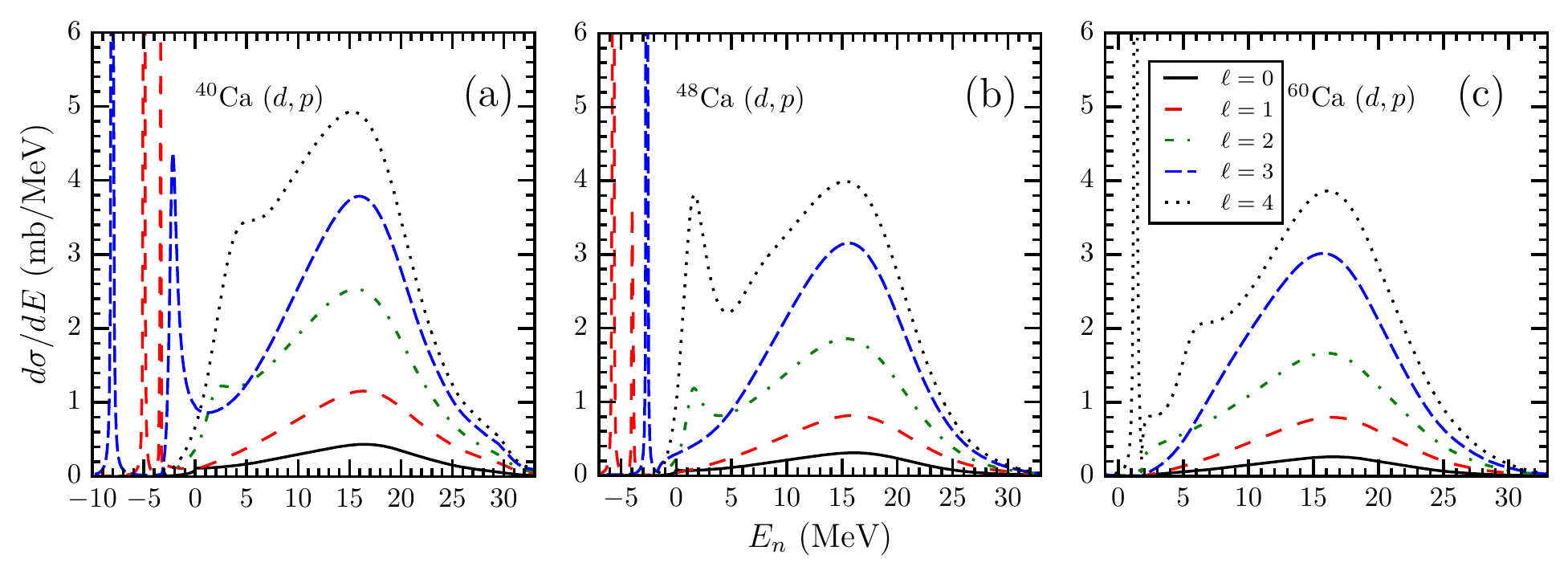}
}
\caption{Population of different final neutron angular momentum states in NEB processes for the reactions $^{40}$Ca$(d,p)$, $^{48}$Ca$(d,p)$, and $^{60}$Ca$(d,p)$. The beam energy is $E_d=40$ MeV.}
	\label{fig_nebl}
\end{figure*}

We next analyze the angular momentum transfer involved in these processes. Because the initial deuteron state is, in our model, $\ell=0$ only, the final angular momentum of the neutron corresponds to the angular momentum transfer in this process. We have considered contributions of up to 9 units of neutron angular momentum, since higher values have been found to be negligible. Most of the contribution to the total EB and NEB cross sections comes from $\ell=0$ to 4. In Fig.\ref{fig_ebl} we show the angular momentum decomposition for the elastic component of (a) $^{40}$Ca$(d,p)$, (b) $^{48}$Ca$(d,p)$ and (c) $^{60}$Ca$(d,p)$ at $E_d=40$ MeV. The contribution from $\ell=0$  (solid black) is small for all cases. The contribution from $\ell=1$ (red dashed) is similar for the three cases. The $\ell=2$ distribution (dot--dashed green) suggests the existence of a resonance at low energy for $^{40}$Ca$(d,p)$, that becomes more pronounced for $^{48}$Ca$(d,p)$ and eventually becomes a strong sharp peak for $^{60}$Ca$(d,p)$. The broad $\ell=3$ peak (long--dashed blue) becomes stronger as one moves away from the valley of stability. Finally, the broad $\ell=4$ contribution (dotted black) is the dominant contribution for all targets at energies $E_n>8$ MeV.

Fig. \ref{fig_nebl} is the same as Fig. \ref{fig_ebl} but now for the non-elastic breakup. We find that the main contributions also come from $\ell=0,1,2,3,4$. The narrow peaks below threshold reflect specific bound states in the $A+1$ system. The DOM predicts several bound states for $^{41}$Ca and $^{49}$Ca which have a correspondence to the experimental known spectrum. Above threshold, the bumps around 2--3 MeV observed for $^{40}$Ca (a), correspond to  a $g_{9/2}$ and  a $d_{5/2}$ resonance. These two peaks move down in energy for $^{48}$Ca and the $g_{9/2}$ state becomes a sharp narrow resonance just above threshold for $^{61}$Ca. DOM thus predicts that $^{61}$Ca is unbound by $E<1$ MeV.
Current ab--initio methods do not have the accuracy to predict whether $^{61}$Ca is bound or unbound.

Other partial waves that contribute significantly to the total cross section are $\ell=1$ and $\ell=3$. The energy distributions are broad and generated by phase space effects in some cases, while in some other cases they reflect the concentration of strength of a given spin and parity over an energy region. Examples of the latter are the $f_{5/2}$ strength peaked around $-2$ MeV in $^{41}$Ca (see Fig. \ref{fig_nebl} (a)) and the $g_{9/2}$ broad resonance in $^{49}$Ca around 1 MeV (see Fig. \ref{fig_nebl} (b)). In both cases, experiments show a number of resolved peaks of the same spin and parity distributed in the same energy region.  

\begin{figure*}
	\centerline {
		\includegraphics*[width=16cm]{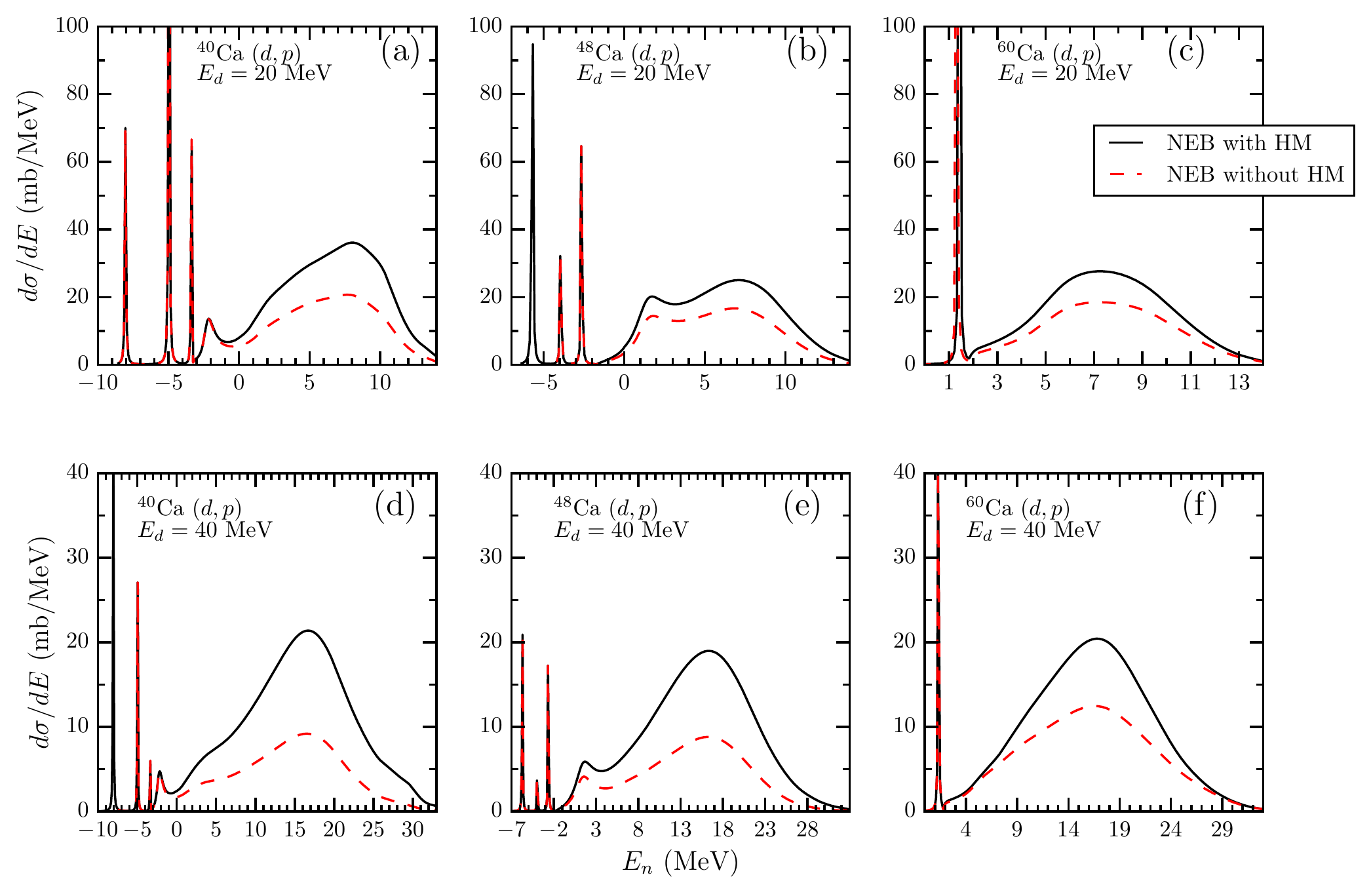}
	}
\caption{Effect of the inclusion of the HM term in the NEB calculations, for the reactions $^{40}$Ca$(d,p)$, $^{48}$Ca$(d,p)$, and $^{60}$Ca$(d,p)$, and for beam energies $E_d=20$ MeV and $E_d=40$ MeV.}
	\label{fig_hm}
\end{figure*}
One important aspect to consider is the effect of the non-orthogonality term, denoted here as HM after Hussein and McVoy \cite{Hussein:85}.
In Fig. \ref{fig_hm} we show the non--elastic breakup cross section for $(d,p)$	 with (black solid lines) and without (red dashed lines) this term for our three targets at $E_d=20$ MeV (top panels) and $E_d=40$ MeV (bottom panels).  It is immediately apparent from these figures that the HM contribution is most significant at the higher energy, even though it does contribute with more than $10$\% for the lower beam energies. The effects at $E_d=40$ MeV can be as large as $100$\%. As far as its dependence on the final neutron energy, this term contributes the most at the peak of the distribution in the continuum, roughly $16$ MeV for the reactions at $E_d=40$ MeV. Its significance becomes smaller for narrow states and can be shown to tend to zero when the width tends to zero (sharp bound states) \cite{Potel:15b}.
 
\begin{figure*}
\centerline {
\includegraphics*[width=16cm]{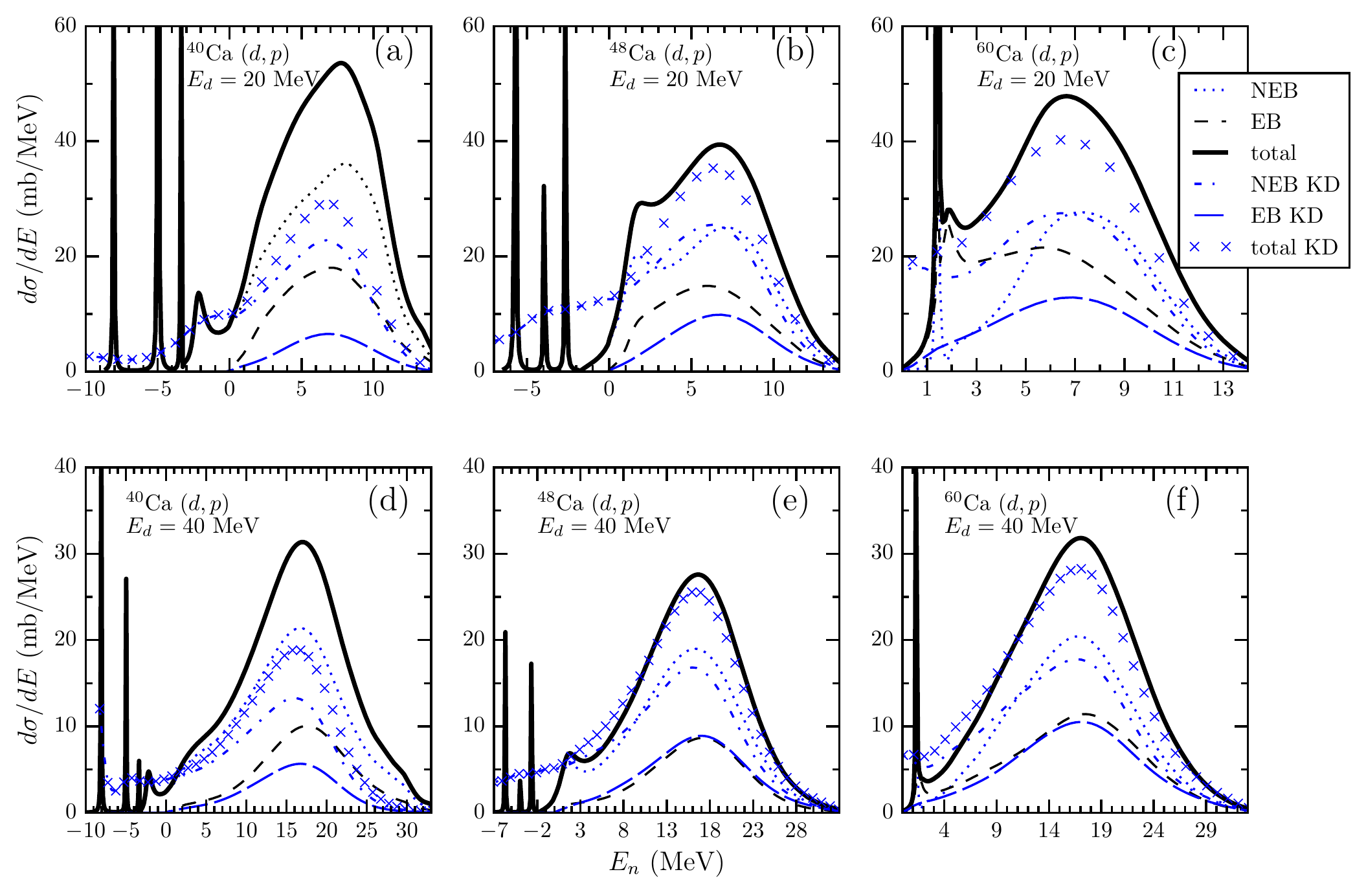}
}
\caption{Comparison of KD phenomenological optical potential and the DOM: elastic breakup (EB) and non--elastic breakup (NEB) proton spectra for the reactions $^{40}$Ca$(d,p)$, $^{48}$Ca$(d,p)$, and $^{60}$Ca$(d,p)$, at $E_d=20$ MeV and $E_d=40$ MeV. }
\label{fig_kd}
\end{figure*}

Although in this work we use the DOM interaction for the calculations, phenomenological potentials that fit scattering observables only and do not connect to bound state properties have widespread use. One recent parametrization is the  KD \cite{Koning:03}. In Fig. \ref{fig_kd} we compare results for elastic breakup (dashed lines) and non--elastic breakup (dotted and dot--dashed lines) resulting from $(d,p)$ reactions, when using the global KD potential  and DOM for the nucleon--target interactions. Also included are the total cross sections (solid lines and crosses). Although KD was not fit to bound states, here we have arbitrarily extended it at negative energies. For physical reasons, and because KD was not fit to negative energies, we impose that the depth of its imaginary terms be  $W_v,W_s>0$ MeV. Not surprisingly the KD predictions differ from the DOM predictions, both above and below threshold. This difference is largest for $^{40}$Ca  where the transfer cross section populating neutron states in the continuum can differ by more than $30$\%. It is to note that a specific version of the KD exists for $^{40}$Ca, fitted on data involving this isotope alone. However, in the calculations presented here we use the global parametrization for all the Ca isotopes. Striking differences are seen also close to threshold, where DOM predicts specific narrow states while KD predicts a smooth background behavior reflecting its lack of structure content.

\begin{figure*}
\centerline {
\includegraphics*[width=16cm]{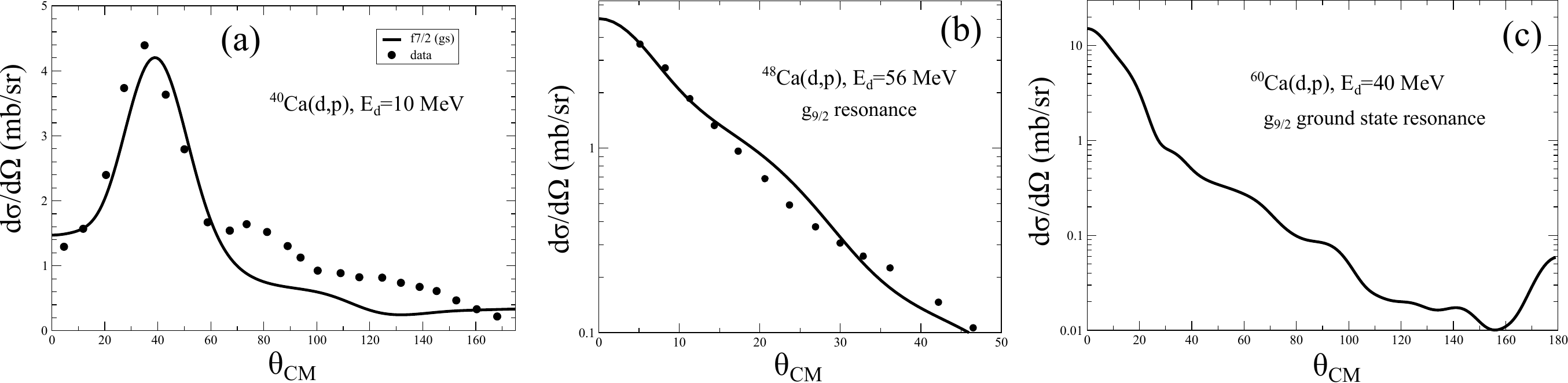}
}
\caption{Angular distributions for: (a) $^{40}$Ca$(d,p)^{41}$Ca(g.s.) at $E_d=10$ MeV \cite{Brown:74},  (b) $^{48}$Ca$(d,p)^{49}$Ca$(9/2^+)$ at $E_d=56$ MeV, and (c) $^{60}$Ca$(d,p)^{61}$Ca(g.s.) at $E_d=40$ \cite{Uozumi:94} MeV. Predictions are compared to data when available.}
\label{fig_ang}
\end{figure*}

Most often, when performing $(d,p)$ measurements,  in addition to the proton (or neutron) energy distributions, one can also obtain the angular distributions to specific final states. In order to ensure that our predictions  indeed provide realistic angular distributions and cross sections, we compare the theoretical results with data for a few cases. In Fig. \ref{fig_ang} (a), we plot the differential cross sections for 
$^{40}$Ca$(d,p)^{41}$Ca at $E_d=10$ MeV, populating the ground state ($f_{7/2}$), compared to data from \cite{Brown:74}. We also show in Fig. \ref{fig_ang}(b) the angular distribution for  $^{48}$Ca$(d,p)^{49}$Ca populating the $9/2^+$ resonance at $E_d=56$ MeV \cite{Uozumi:94}, and in Fig.\ref{fig_ang}(c), the prediction for $^{60}$Ca$(d,p)^{61}$Ca populating the predicted g.s. resonance. 
Note that no free parameters have been introduced in this description. In particular, the states predicted by DOM are not exactly at the physical energies. Also, since the DOM already takes into account many-body structure effects, we do not scale our predictions by any spectroscopic factor. 

Fig. \ref{fig_ang} demonstrates that the shapes of the angular distributions are fairly well reproduced in our model. For the case in which the transfer reaction populates a bound state ($^{40}$Ca$(d,p)^{41}$Ca(g.s.)), the  magnitude is well predicted by DOM, as expected from previous studies \cite{Nguyen:2011}.
In Fig. \ref{fig_ang} we also show angular distributions that populate neutron states in the continuum. In order to compare with experiment, we need to integrate these differential cross sections over the given neutron energy bin. In the $^{49}$Ca$(9/2^+)$ this is not trivial because the resonance is broad and sitting on a background from other partial waves. Thus, for this case, we take the angular distribution for $E_n=1.7$ MeV (the peak of the resonance) and then arbitrarily scale the cross section to match the experimental value. When  analyzing the data, one would have to use  the exact experimental binning for a meaningful quantitative interpretation. The shape of the angular distribution predicted by our model matches the measured angular distribution. 
Finally, we also present our predictions for $^{61}$Ca populating the predicted $9/2^+$ resonance at $E_d=40$ MeV.

%% file: Results_gamma_decay.tex
% J. Escher - 2017-02-22

% \providecommand{\lambdabar}{\mathchar'26\mkern-10mu\lambda}

\subsection{Application to $(n,\gamma)$ surrogates}
\begin{figure*}
\centerline {
\includegraphics*[width=6cm,trim=0 0 60 70,clip=true]{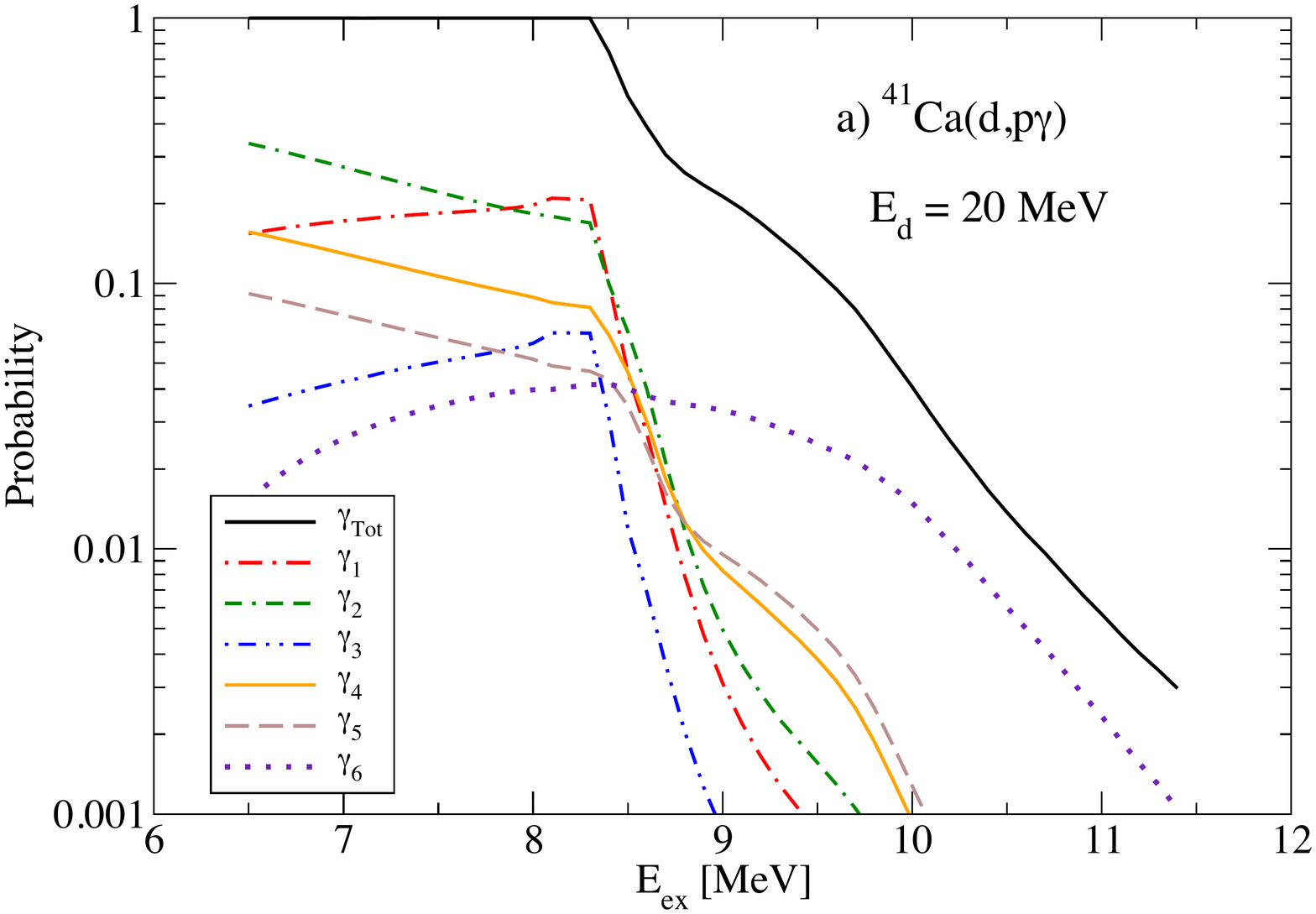}
\includegraphics*[width=6cm,trim=0 0 60 70,clip=true]{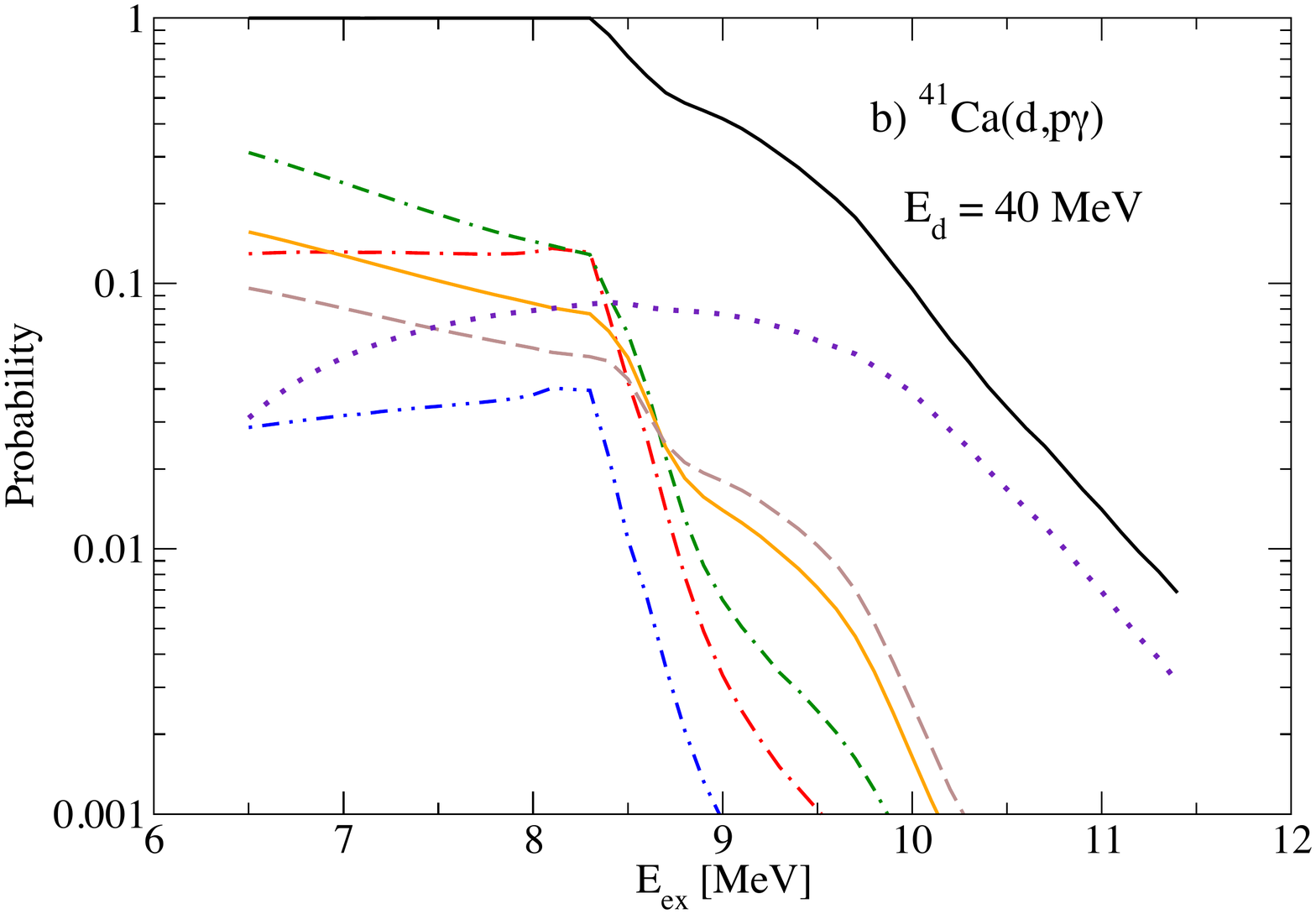}
\includegraphics*[width=6cm,trim=0 0 60 70,clip=true]{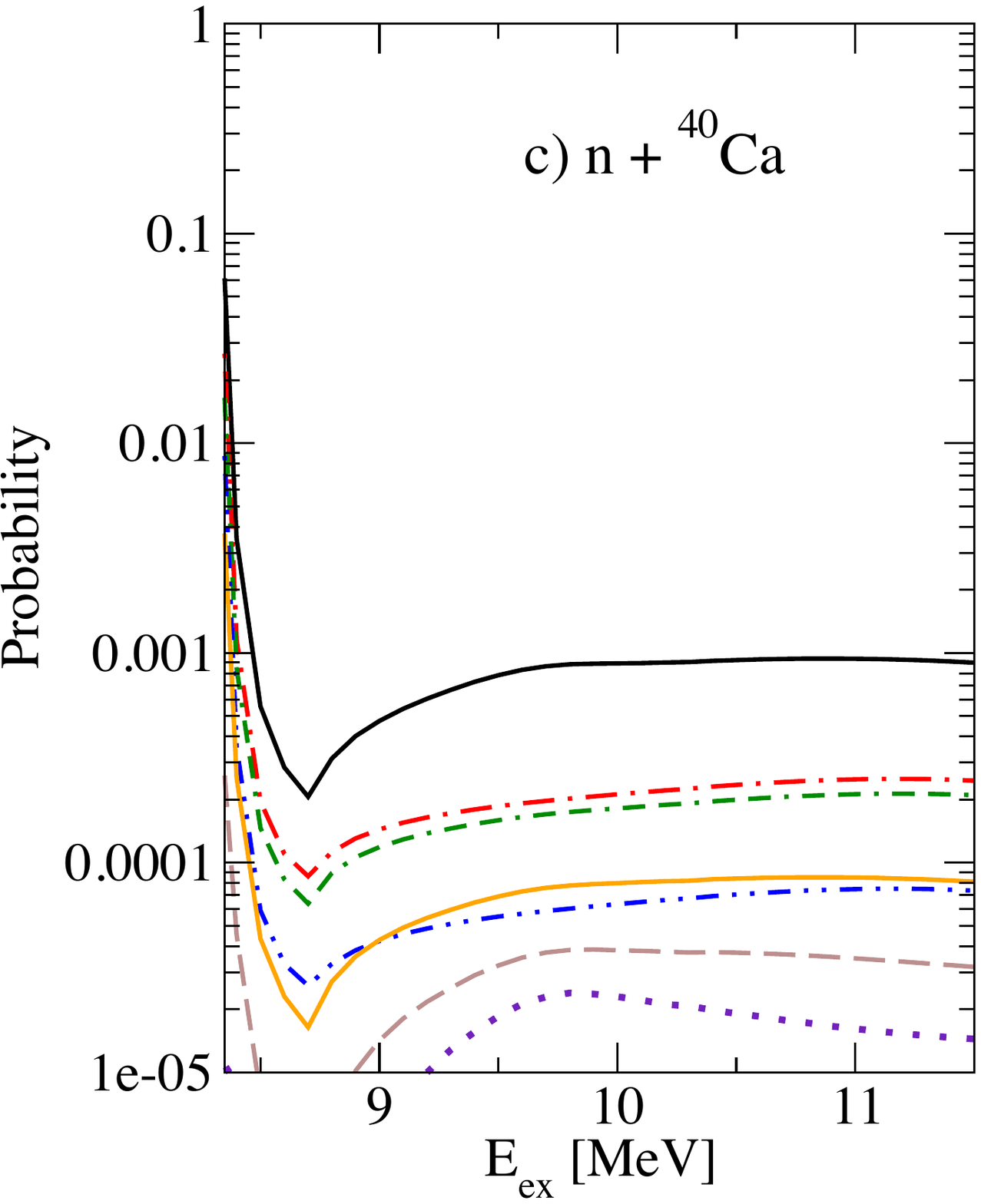}
}
\centerline {
\includegraphics*[width=6cm,trim=0 0 60 90,clip=true]{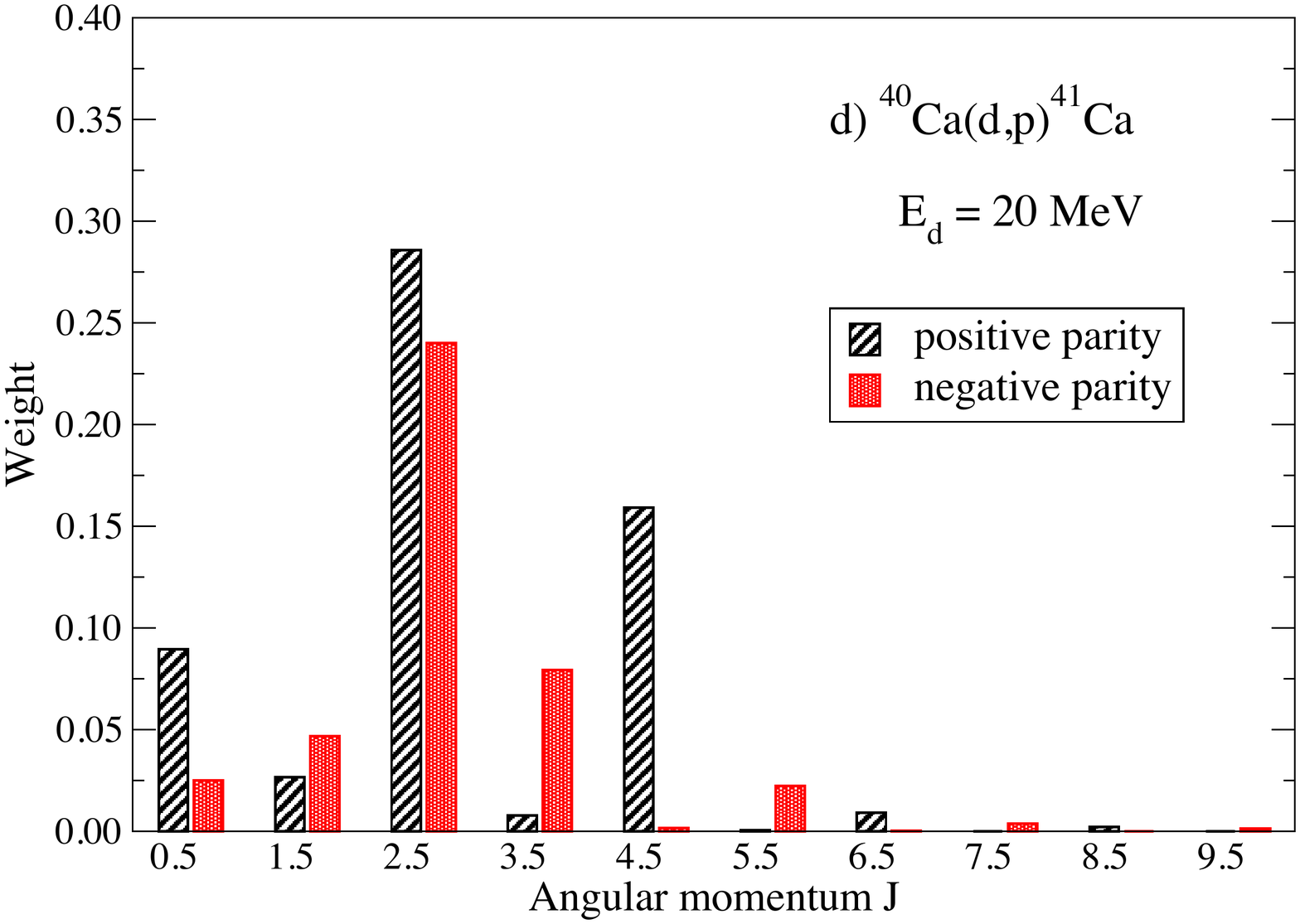}
\includegraphics*[width=6cm,trim=0 0 60 90,clip=true]{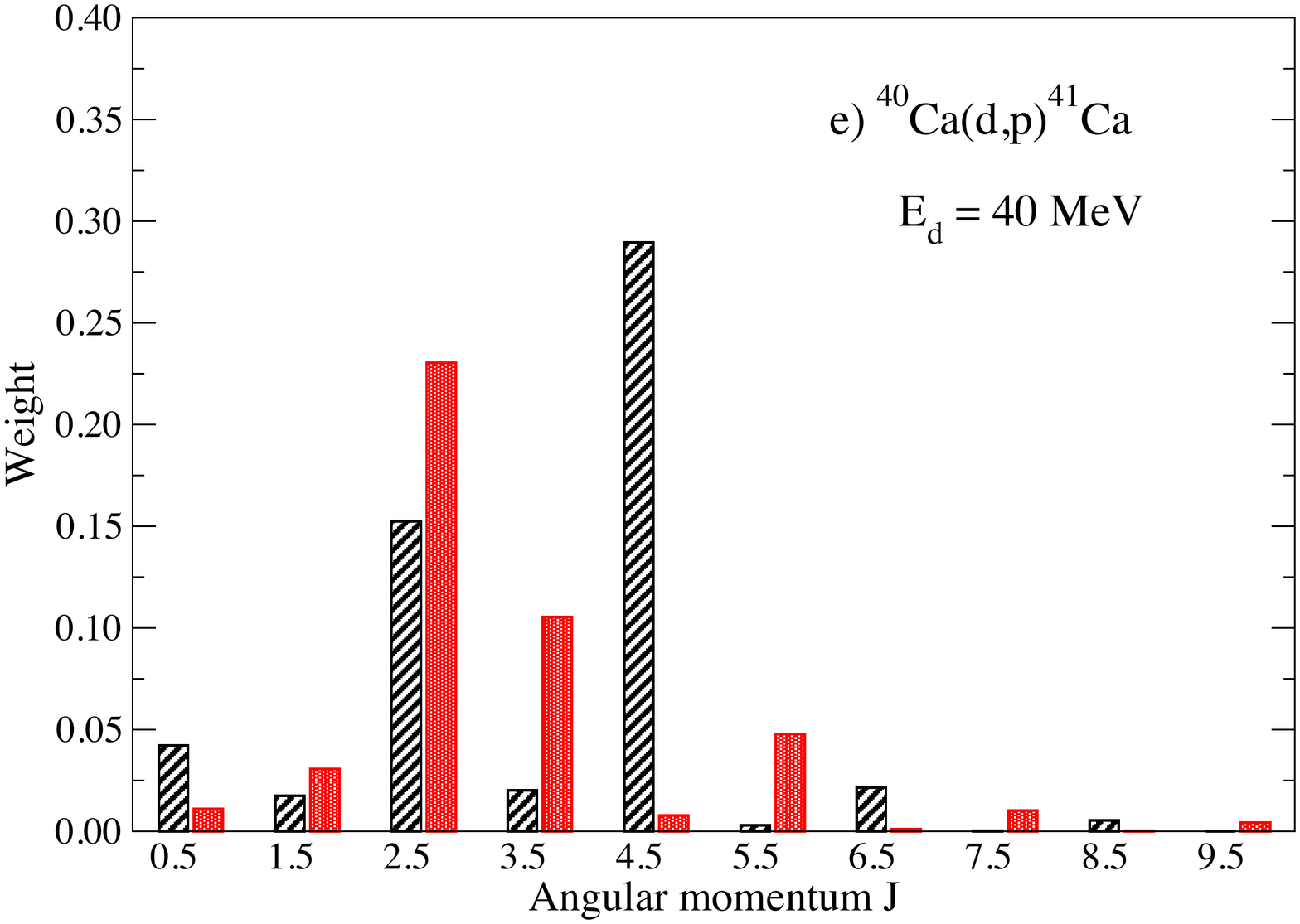}
\includegraphics*[width=6cm,trim=0 0 60 90,clip=true]{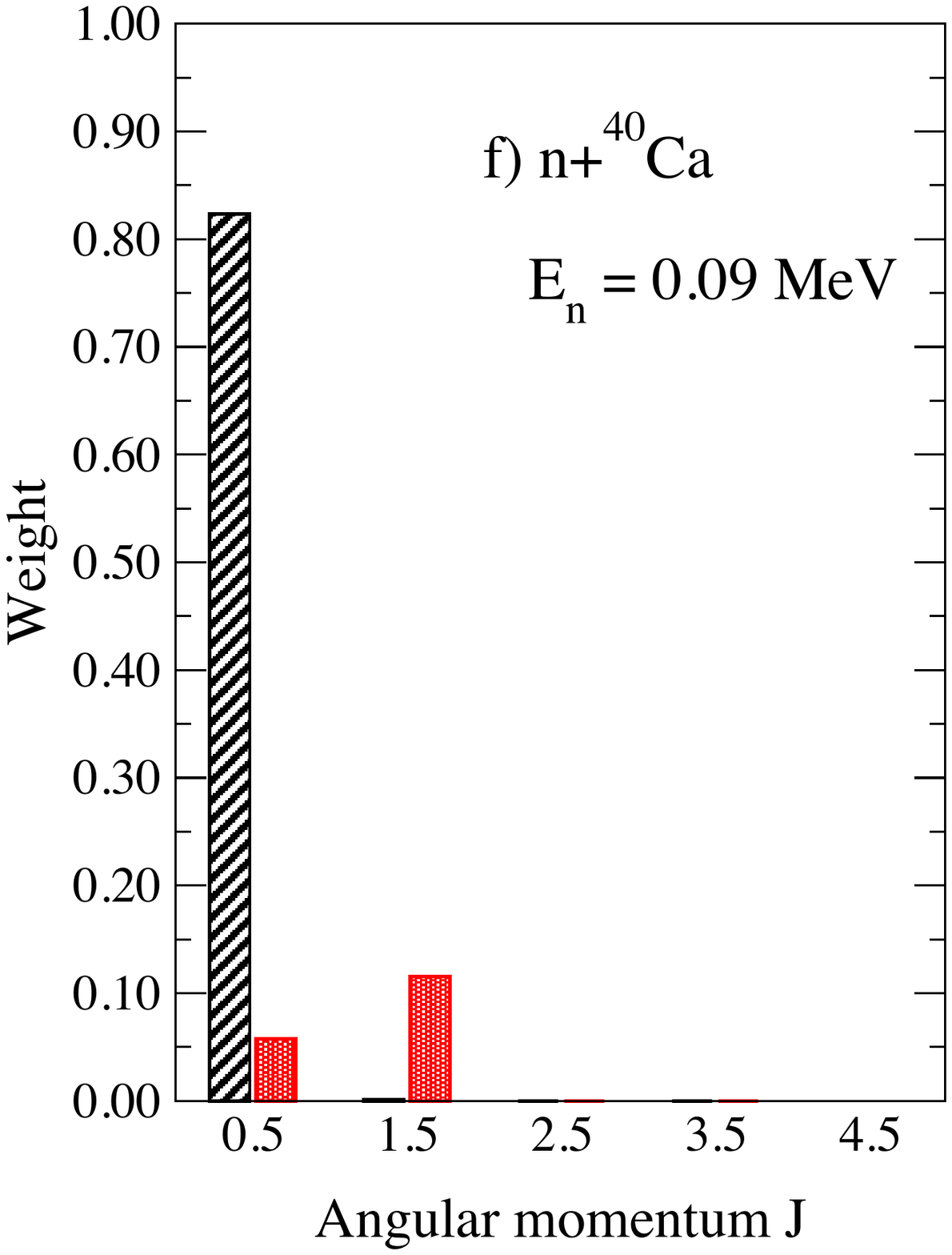}
}
\caption{Decay of the compound nucleus $^{41}$Ca following (a) $^{40}$Ca$(d,p)$ at $E_d$ = 20 MeV (b)  $^{40}$Ca$(d,p)$ at $E_d$ = 40 MeV (c) neutron capture on  $^{40}$Ca.  Shown are the probabilities for observing six selected $\gamma$ transitions in  $^{40}$Ca, as a function of the excitation energy $E_{ex}$ at which this nucleus was populated.  Also shown is the total probability for decay into the $\gamma$-channel (solid black curve).  The bottom panels show the associated spin-parity distributions at $E_{ex} = 8.44$ MeV, i.e. a little less than 100 keV above the neutron separation energy of $S_n = 8.36$ MeV.}
\label{fig_41CaDecay}
\end{figure*}

The angular-momentum transfer involved in the $(d,p)$ reactions discussed here differs significantly from the angular-momentum transfer that occurs in a typical low-energy neutron-induced reaction. Figure \ref{fig_ebneb} shows that the non-elastic breakup processes include strong contributions for transfers up to $\ell=4$. Consequently, the decay pattern of the resulting compound nucleus will differ from that observed in a neutron capture reaction.  This is illustrated in Figure~\ref{fig_41CaDecay}, which compares the expected $\gamma$-decays of $^{41}$Ca following $^{41}$Ca$(d,p)$, for two different energies $E_d$ = 20 MeV and 40 MeV, to each other and to the decay following neutron capture with $E_n < 4$ MeV.

In the top panels, we show the probabilities for observing specific $\gamma$-ray transitions in the decaying $^{41}$Ca when this nucleus is populated at excitation energy $E_{ex}$. 
Specifically, we consider 
$\gamma_1$: $3/2^-$(1.94 MeV) $\rightarrow 7/2^-$(g.s.), 
$\gamma_2$: $3/2^+$(2.01 MeV) $\rightarrow 7/2^-$(g.s.), 
$\gamma_3$: $3/2^-$(2.46 MeV) $\rightarrow 3/2^-$(1.94 MeV), 
$\gamma_4$: $5/2^+$(2.61 MeV) $\rightarrow 7/2^-$(g.s.), 
$\gamma_5$: $7/2^+$(2.88 MeV) $\rightarrow 7/2^-$(g.s.), 
$\gamma_6$: $9/2^-$(3.20 MeV) $\rightarrow 7/2^-$(g.s.).
Such probabilities have been observed in experiments in which the ejectile of a transfer reaction (here the proton) is detected in coincidence with a characteristic $\gamma$-transition, see, e.g. Refs.~\cite{Ota:15b,Scielzo:10a,Hatarik:10a}.  For the $(d,p)$ transfer reactions we give probabilities for energies below as well as above the neutron separation energy, $S_n = 8.36$ MeV, while an incident neutron can only populate states above that.  In all cases, we also show that total probability for decay into the $\gamma$-channel (solid black curve).

The calculations were carried out by combining the predicted spin-parity distributions from the $^{41}$Ca$(d,p)$ calculations described above with a Hauser-Feshbach-type decay model.  The parameters of the decay model were taken from recent evaluations of n+$^{40}$Ca cross sections~\cite{Capote:09,Shibata:11,Chadwick:11}.  Specifically, the $\gamma$-ray strength function is dominated by an E1 component with parameters taken from the Reference Input Parameter Library RIPL-3~\cite{Capote:09} and includes also smaller M1 and E2 contributions. For the level density, a Gilbert-Cameron prescription was adopted, with parameters taken from the JENDL 4.0 evaluation~\cite{Shibata:11}. Slight adjustments were made to reproduce available $(n,\gamma)$, $(n,p)$, and $(n,\alpha)$ cross sections.

%The calculations were carried out by combining the predicted spin-parity distributions from the $^{41}$Ca(d,p) calculations described above with a Hauser-Feshbach-type decay model.  The parameters of the decay model were taken from recent evaluations of n+$^{40}$Ca cross sections~\cite{Capote:09,JENDL40_Ca,ENDFb7_Ca} and slightly adjusted to reproduce available $(n,\gamma)$, $(n,p)$, and $(n,\alpha)$ cross sections.

We observe that all transitions have probabilities of 30\% or less, which is common for decays of odd nuclei.  
Increasing the deuteron energy from 20 MeV to 40 MeV moves the compound-nucleus spin-parity distribution to larger angular momenta, as is shown for $E_{ex}$ = 8.44 MeV in panels (d) and (e).  With this shift $\gamma$--ray transitions $J_i \rightarrow J_f$ that involve larger $J_i$ are enhanced relative to those with smaller $J_i$.   This can be clearly seen for the $\gamma_6$ (9/2$^+$ $\rightarrow$ 7/2$^-$) transition. Electric dipole transitions dominate the $\gamma$ cascade, thus a compound nucleus with a significant high angular--momentum population decays predominantly through transitions with large $J_i$.

The strong drop-off of the $\gamma$ probabilities at $E_{ex} = S_n =$ 8.36 MeV is due to the opening of the neutron channel.
The decline is particularly strong for those transitions that are fed by the low angular--momentum portion of the CN spin-parity distribution.  Neutron emission to the $0^+$ ground state of $^{40}$Ca competes strongly in those cases, while it is hindered for the high-angular momentum portion of the CN spin-parity distribution. A striking example is given by the $\gamma_6$ which exhibits almost no drop-off until the excitation energy is increased to more than 1 MeV above $S_n$ and d-wave neutron emission becomes possible, in addition to s- and p-wave emission.  

For the neutron-induced reaction, shown in panels (c) and (f), the compound-nuclear decay looks quite different. The CN spins are quite low and, consequently, the $\gamma$--probabilities drop strongly as the neutron emission threshold is crossed. Overall, the $\gamma$ probabilities are very low, as neutron emission to the $^{40}$Ca ground state involves small changes in angular momentum and is thus strong at all energies.

These observations make it clear that a $(d,p\gamma)$ reaction cannot be simply used as a stand-in for an $(n,\gamma$) reaction, but that additional work is required to extract a neutron capture cross section from a $(d,p)$ one-neutron transfer experiment.  In particular, a careful accounting of spin effects is required.  The method outlined in Section~\ref{surrogates}, which aims at constraining the Hauser--Feshbach inputs by fitting the parameters in the decay model so that measured $\gamma$ probabilities are reproduced, is being explored and shows promise~\cite{Ratkiewicz:15,Ratkiewicz:17,Escher:17a}.
On the other hand, measurements of the behavior of such $\gamma$--ray probabilities can be used to test theoretical descriptions of deuteron--induced reactions in cases where sufficient independent information on the $\gamma$-cascade of the decaying nucleus is available. A comprehensive set of $(d,p\gamma)$ cross section data for $^{40}$Ca would be needed to explicitly apply the method in the present case.

%For instance, the probability for the $\gamma_6$ (9/2$^+$ $\rightarrow$ 7/2$^-$) transition, which involves the population of states with larger angular momenta, is enhanced in the 40-MeV (d,p) reaction, relative to the 20-MeV (d,p) reaction, and is strongly suppressed in the n-induced reaction.

%% file: Discussion.tex
We  review  the status of inclusive deuteron--induced reactions, benchmarking three recently developed codes (\cite{Lei:15,Carlson:15,Potel:15b}), allowing for the breakup--fusion formalism to rest on firm theoretical and numerical grounds. We point out the importance of including the HM term, showing explicitly its contribution in the $^{40,48,60}$Ca $(d,p)$ calculations.  We compare the finite--range calculation with  the zero--range approximation, confirming the good agreement found in a very recent publication \cite{Lei:17}.  A dispersive optical model for the neutron--target optical potential (see \cite{Dickhoff17} and refs. therein) is used, comparing some of the calculations with the results obtained using a global parametrization (Koning--Delaroche, \cite{Koning:03}). Finally, we present applications in nuclear  structure (both below and above the neutron emission threshold), and the description of the formation of compound nucleus states with a given energy, spin, and parity. We show explicitly that the latter is essential for the use of $(d,p\gamma)$ reactions as surrogates to $(n,\gamma)$ processes.

We compute differential cross sections, and show that the energy position of  the principal single--particle--like peaks is reproduced within about 500 keV, and the absolute value of the cross section gives information about the distribution of strength across the full energy spectrum. Spectroscopic factors, associated with the contribution of a single particle orbit defined in an arbitrary basis, are not used in our framework. The DOM concentrates this strength in the principal peak for the first few excited states, and smears out the multiple fragments of a given spin and parity around a broad peak for higher excitation energies. Resonances in the continuum are also described,  although individual resonances are usually not resolved. Extrapolation to very exotic isotopes, such as $^{60}$Ca, can be performed assuming a particular dependence on the asymmetry of the optical potential. The accuracy of such predictions still awaits confirmation, but the calculations showcase the capabilities of the formalism. While reaction experiments using radioactive $^{60}$Ca beams may be still far in the future due to low beam intensities, the projected ability of FRIB to nominally produce this isotope means that we might be able to check the predictions of the DOM (a crucial element of the current calculations) even for isotopes near the driplines in the near future. 

We address the application  of $(d,p\gamma)$ reactions as surrogates to $(n,\gamma)$ capture processes. The determination of the spin--parity distribution of the compound nucleus is crucial for the implementation of the method. This distribution depends on details of the neutron--target optical potential, as well as on the deuteron beam energy. Once the compound nucleus created in the $(d,p)$ reaction has been characterized, a Hauser--Feshbach analysis of the observed $\gamma$ decay may allow one to infer the $(n,\gamma)$ sections. We illustrate this method by showing the predicted $\gamma$ decay in a $^{40}$Ca$(d,p\gamma)$ reaction. Our results are shown to be very sensitive to the spin--parity distribution of the compound nucleus populated at the two beam energies considered,  the method presented here is thus very promising in improving the potential of surrogate reactions.  A challenging task for the future is to be able to discriminate the different contributions to the NEB. More specifically, it is important to compute separately the direct, pre--equilibrium and compound processes, at least in energy regions where they are competing. 

Moving towards a complete and reliable theory for deuteron--induced reactions, we identify  some future development areas. One of the main sources of uncertainty in the current form of the reaction formalism is the use of phenomenological optical potentials to describe the incoming deuteron. These potentials are fit to elastic data, and  using them in the context of transfer and breakup reactions is known to give rise to uncontrolled inaccuracies. An important and challenging development would be to describe consistently the entrance channel within the same three--body (neutron--proton--target)  effective Hamiltonian used throughout.

Although the DOM has provided  a good account of the essential properties of the nucleon--target optical potential, it is important to note that only its nonlocal implementation can give a complete account of ground--state properties \cite{Dickhoff17}, as well as provide more appropriate distorted waves in the interior of the nucleus. In addition, future developments should  also focus on structure aspects that can be probed with the $(p,d)$ reaction, in keeping with the fact that this probe allows for a different set of states to be populated. Also relevant, progress with microscopical, \textit{ab--initio} as well as semi--phenomenological (e.g. Nuclear Field Theory (NFT), Random Phase Approximation (RPA) or Quasi--Particle Random Phase Approximation (QRPA)) calculations of the self--energy could be incorporated in the formalism. Within this context,  the construction of microscopical optical potentials within the Coupled Cluster formalism has been recently reported \cite{Rotureau:17}. In order to profit from these developments,  one would have  to adapt the codes for nonlocal self--energies, with an arbitrary dependence on energy, spin, and parity.

Now that the $(d,p)$ inclusive breakup theory is proven to be on solid formal and computational grounds, we envisage to generalize the theory to more complicated projectiles containing three fragments, such as the triton. The study of inclusive proton spectra in a $(t,p)$ reaction would then be possible and the investigation of the mechanism of $2n$ capture would be possible. Such a theory, which is a natural extension of the IAV one, has recently been developed in \cite{Carlson:17}, and numerical calculation of the cross section is currently in progress.

Finally, we stress the important  synergy between theory and experiment. The description of the variety of different channels populated in a deuteron--induced reaction within a unique consistent framework enables the comparison with integral measurements along isotopic chains, and over a wide energy range. The predictions could then be tested, and the reliability of the theory improved, in view of the  expected availability of new exotic isotopes in FRIB.
\begin{acknowledgement}
We gratefully acknowledge H. Crawford for useful comments and discussions. This work results from the International Collaboration in Nuclear Theory (ICNT) program at NSCL/FRIB in 2016. We thank the NSCL/FRIB for the logistic and financial support that made this program possible.
This work is performed in part under the auspices of the DOE by Lawrence Livermore National Laboratory under Contract No. DE-AC52-07NA27344, with funding provided through the LDRD project 16-ERD-022. 
This work was supported by the U.S. National Science Foundation under 
grants PHY--1613362, PHY--1403906, PHY--1520929, and by the U.S. Department of Energy NNSA under Contract No. DE--FG52--08NA28552. 
We thank the Institute for Nuclear Theory at the University of Washington for its hospitality and the Department of Energy for partial support during the completion of this work.
\end{acknowledgement}